\documentclass[manuscript=article]{achemso}

\usepackage[version=3]{mhchem} 
\usepackage{gensymb}
\usepackage{siunitx}
\usepackage{hyperref}

\author{Simon W. Briesenick}
\affiliation[Physics Department]{Department of Physics, McGill University, Montr{\'e}al, Canada}
\altaffiliation{These authors contributed equally to this work.}

\author{Wyatt A. Behn}
\email{wyatt.behn@mail.mcgill.ca}
\affiliation[Physics Department]{Department of Physics, McGill University, Montr{\'e}al, Canada}
\altaffiliation{These authors contributed equally to this work.}

\author{Manuel Gonz{\'a}lez Lastre}
\email{manuele.gonzalez@uam.es}
\affiliation{Departamento de F{\'i}sica Te{\'o}rica de la Materia Condensada, Universidad Aut{\'o}noma de Madrid, E-28049 Madrid, Spain}
\altaffiliation{These authors contributed equally to this work.}

\author{Chang Wan Kang}
\affiliation[Chemistry Department]{Department of Chemistry, McGill University, Montr{\'e}al, Canada}

\author{Pablo Pou}
\affiliation{Departamento de F{\'i}sica Te{\'o}rica de la Materia Condensada, Universidad Aut{\'o}noma de Madrid, E-28049 Madrid, Spain}
\altaffiliation{Condensed Matter Physics Center (IFIMAC), Universidad Aut{\'o}noma de Madrid, E-28049 Madrid, Spain}

\author{Ekaterina D. Ulyanov}
\affiliation[Physics Department]{Department of Physics, McGill University, Montr{\'e}al, Canada}

\author{Rub{\'e}n P{\'e}rez}
\affiliation{Departamento de F{\'i}sica Te{\'o}rica de la Materia Condensada, Universidad Aut{\'o}noma de Madrid, E-28049 Madrid, Spain}
\altaffiliation{Condensed Matter Physics Center (IFIMAC), Universidad Aut{\'o}noma de Madrid, E-28049 Madrid, Spain}

\author{Dmytro F. Perepichka}
\email{dmytro.perepichka@mcgill.ca}
\affiliation[Chemistry Department]{Department of Chemistry, McGill University, Montr{\'e}al, Canada}

\author{Peter Grutter}
\email{peter.grutter@mcgill.ca}
\affiliation[Physics Department]{Department of Physics, McGill University, Montr{\'e}al, Canada}

\title[Paper Title]
{Mapping the Growth of Two-Dimensional $\pi$-Conjugated Polymers on Au(111): Organometallic Intermediates and Edge Terminations}

\abbreviations{}

\begin{document}

\begin{abstract}
Kagome lattices provide an exciting space for the exploration of graphene-like $\pi$-conjugated molecular systems with flat bands. Using heterotriangulene-derived precursors along with an on-surface Ullmann coupling process makes growing polymers with Kagome lattices accessible and straightforward. Here, we use scanning tunneling microscopy alongside high-resolution atomic force microscopy to examine the evolution of tribromotrioxaazatriangulene on Au(111) into ordered, covalent films. Using density functional theory and scanning probe methods, we find previously unreported organometallic intermediate states involving Au adatoms incorporated within the growing polymer lattice. We also find that a majority of polymer edges remain brominated up to 250 $^{\circ}$C and a large number of edges bonded to Au adatoms coordinated to an adjacent bromine atom. These observations suggest that residual bromine could play a role in stabilizing the polymer edges to Au adatoms and thereby influence the growth pathways that lead to ordered Kagome polymer lattices.
\end{abstract}

\section{Introduction}
Two-dimensional (2D) materials with Kagome lattices have attracted significant interest due to their electronic structure, which features both Dirac cones with massless charge carriers and flat bands favorable for correlated electron physics~\cite{wang2024topological}. 2D conjugated polymers (2DCPs) based on triangulene-derived monomers are a versatile materials platform for realizing Kagome lattices, particularly since the electronic, structural, and optical properties may be tuned to a large degree by modification of the chemical building blocks~\cite{bieri2011surface,schlutter2013pi,steiner2017hierarchical,galeotti2020synthesis,ni2022organic,dettmann2023electronic,briesenick2024kinetic,pawlak2025surface}. For practical applications, controllable and reproducible synthesis of well-defined 2DCPs is essential. However, the production of ordered semiconducting 2DCP films, such as those produced through surface-assisted Ullmann coupling, continues to be limited by topological defects, grain boundaries, and residual reaction products~\cite{grill2007nano,grill2020covalent,qin2024recent}. Efforts to mitigate disorder have included optimization of growth temperature and deposition rate~\cite{eichhorn2014influence}, selection of metallic substrate~\cite{bieri2010two}, variation of the terminating halogens~\cite{galeotti2017role}, and hierarchical growth schemes~\cite{lafferentz2012controlling}. In spite of this, realistic single-grain sizes are still limited to the nanoscale. Furthermore, the interplay between kinetically and thermodynamically controlled reaction steps complicates the predictability of more ideal structures, creating additional barriers to achieving higher-quality 2DCPs~\cite{fritton2019role}. 

Given these challenges, the importance of mapping a given precursor's growth parameter space and addressing its realistic, long-range crystalline limitations must be emphasized, as all possible intermediate structures and all reaction pathways are not always known \textit{a priori}~\cite{lischka2018surface,galeotti2019unexpected}. For 2DCPs synthesized via surface-assisted Ullmann coupling, the polymer lattice symmetry, dehalogenation temperature, covalent coupling rate, and rigidity of the precursor will all have notable effects on the final product~\cite{eichhorn2014influence,lackinger2017surface}. 

To probe this reaction landscape, we explored the on-surface growth of tribromotrioxaazatriangulene (TBTANG) on Au(111) as a function of sample temperature and precursor deposition rate. TBTANG is chosen as it is a chemically stable precursor with a closed-shell electronic structure which forms a highly-ordered $\pi$-conjugated framework~\cite{galeotti2020synthesis}. With a combination of high-resolution scanning tunneling microscopy (STM), CO-functionalized non-contact atomic force microscopy (nc-AFM), and density functional theory (DFT) calculations, we directly identify the local chemical states across several stages of the reaction pathway, with a focus on the edge states, up to the covalently-coupled polymer.

We noted the presence of organometallic (OM) intermediates featuring dehalogenated radicals coordinated to Au adatoms. These adatoms are most likely provided by the Au surface edges, and previous literature found a correlation between the coinage metal adatom population and the presence of adsorbates, suggesting that the chemisorption step of the Ullmann coupling process provides a more energy-favorable environment for the formation of surface adatoms~\cite{barton2017formation}. These OM structures are less prevalent on Au than on other metal surfaces, particularly Ag and Cu~\cite{eichhorn2014intermediate,wang2023metal,herrera2025surface,chen2025density,xu2023formation}. At low surface annealing temperatures, we observe a predominantly metal-coordinated network over a covalently-coupled one, which indicates a high coupling energy barrier $E_{\text{c}}$ relative to debromination $E_{\text{dehal}}$. As the annealing temperature is increased, the density of OM intermediates sharply drops, and the quality of the covalent network improves. While prior work on heterotriangulene precursors on Au(111) has focused on characterizing covalently coupled networks, the transient Au-adatom-coordinated OM intermediate state has, to the authors' knowledge, neither been observed nor explicitly modeled~\cite{bieri2011surface,steiner2017hierarchical,galeotti2020synthesis,dettmann2023electronic}. Additionally, we observed a persistence of surface-bound bromine after annealing up to 250~\si{\celsius} as well as polymer edges largely terminated with bromine rather than stabilized by Au surface atoms or adatoms, which points to a gradual dehalogenation process. We go on to classify and chemically identify the other predominant edge terminations of the polymer as a function of anneal temperature. Robust agreement between experimental observations and simulated molecular geometries supports our structural assignments of polymer edge terminations and OM intermediates. 

\section{Results and Discussion}
\subsection{Polymer Characterization}
\begin{figure}
    \centering
    \includegraphics[width=\linewidth]{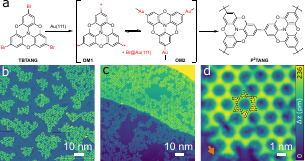}
    \caption{P$^2$TANG polymer formation and imaging. (a) Polymerization reaction of TBTANG into P\textsuperscript{2}TANG. (b,c) 100 nm STM images of the growing polymer. (d) 10 nm STM image with molecular overlay. The arrow points to a bromine atom. Imaging parameters: (b) $V_s$ = -500 mV, $I_t$ = 50 pA; (c) -300 mV, 50 pA; (d) 300 mV, 30 pA. Film growth parameters: (b) $T_s$ = 180~\si{\celsius}, $T_c$ = 140~\si{\celsius}, $t_d$ = 1.75 min; (c) 210~\si{\celsius}, 140~\si{\celsius}, 3.5 min; (d) 250~\si{\celsius}, 140~\si{\celsius}, 3.5 min.}
    \label{polymerization_reaction}
\end{figure}

Each polymer growth is categorized by the sample anneal temperature ($T_s$), the temperature of the molecular source crucible ($T_c$), and the exposure time to the molecular flux ($t_d$). The polymerization reaction of brominated TANG monomers into  the corresponding two-dimensional conjugated polymer, P\textsuperscript{2}TANG, is outlined in Figure~\ref{polymerization_reaction}a. TBTANG is terminated with bromine atoms which are cleaved on the hot ($T_s$ = 180--250~\si{\celsius}) \ce{Au}(111) surface, as previously established by X-ray photoelectron spectroscopy (XPS)~\cite{galeotti2020synthesis}. The dehalogenated molecules (formally triradicals) are bound to the underlying surface, either with surface atoms (arrangement OM1) or adatoms (OM2)~\cite{galeotti2019unexpected,hu2020ullmann,zhang2021adatoms,zhao2025unconventional}. OM1 allows for the diffusion of dehalogenated molecules along the surface and their encounter forming the covalent C---C bond~\cite{lackinger2017surface}. OM2 reversibly binds the molecules in the OM network (common on Cu~\cite{zint2017successive,galeotti2019unexpected,grill2020covalent} but also observed on Au~\cite{lischka2018surface,galeotti2019thiophene}, and can serve as a pool for the more reactive OM1~\cite{dettmann2024real}). Figure~\ref{polymerization_reaction}b shows an STM scan taken at low temperature (9.6~K) under ultra-high vacuum (UHV) of P\textsuperscript{2}TANG/\ce{Au}(111) with submonolayer coverage. For this growth at $T_s$ = 180~\si{\celsius}, numerous islands are present and beginning to merge into one another; the spatial distribution of the islands themselves reflects the underlying symmetry of the (111) surface facet, having a triangular superstructure. Probable nucleation sites are the herringbone ``elbow" sites of the well-known $22\times\sqrt{3}$ surface reconstruction~\cite{pham2015heat-induced}. For the same $T_c$ = 140~\si{\celsius}, but this time for $T_s$ = 210~\si{\celsius}, a different nucleation pattern emerges. In Figure~\ref{polymerization_reaction}c, we no longer observe separate islands, but rather polymer patches with small voids and larger bare Au regions.  Figure~\ref{polymerization_reaction}d shows a detailed STM image of a region of the resulting polymer formed at 250~\si{\celsius}. The observed Kagome lattice with a lattice constant of 1.73$\pm$0.06~nm is consistent with earlier results~\cite{galeotti2020synthesis,dettmann2023electronic}; however, defects are evident within. Irreversible C---C bonds ``lock in" vacancies during the growth, as the growing structure sterically hinders monomers from reaching these vacancy positions. Additionally, two equivalent rotational domains prevent the formation of one continuous sheet (Supporting Information Figure~S2). The lack of self-correction and Ostwald ripening mechanisms are well-known limitations of on-surface Ullmann coupling~\cite{grill2020covalent}. 

\begin{figure}
    \centering
    \includegraphics[width=\linewidth]{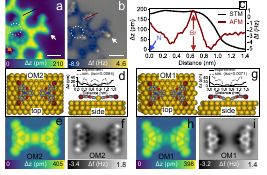}
    \caption{(a,b) Experimental characterization: STM ($V_s$ = -300~mV, $I_t$ = 50~pA) and high-resolution AFM ($f_0$ = 24,304 Hz, $A$ = 55 pm, $Q$ = 190,385) images of a P\textsuperscript{2}TANG polymer edge. Scale bars are 1 nm. Orange arrow, surface-adsorbed bromine atom; white arrows, C$\rm{-}$Br feature; dashed circle, OM intermediate. (c) Edge termination profiles extracted from lines in (a,b). (d,g) Simulated geometries for a partially brominated TANG dimer bonded to an Au adatom (d) and to an Au surface atom (g). Insets compare the experimental OM2 line profile to simulated STM line profiles for OM1 and OM2 intermediates. Profiles are extracted along the center of the intermediate and simulated profiles are scaled 0.5$\times$. (e,h) Simulated constant-current STM contrast (filled density of states $V_s$ = -500 mV) with isosurface (iso) values of 0.043  e/ \AA$^{3}$ (e) and 0.041  e/ \AA$^{3}$ (h). (f,i) FDBM AFM simulations for the respective geometries at $z$ = 700 pm from the surface. Film growth parameters: $T_s$ = 210~\si{\celsius}, $T_c$ = 120~\si{\celsius}.}
    \label{intermediates_summary}
\end{figure}

Bright features, which are around half the apparent height of the polymer in constant-current STM contrast, are observed within the pores and around the edges of the P\textsuperscript{2}TANG in Figure~\ref{polymerization_reaction}d (an example is indicated by an arrow). These features are interpreted as residual bromine which becomes trapped within enclosed gaps during polymerization; the stoichiometry of the reaction predicts 6 halogens per hexagonal pore. Adsorbed bromine atoms are clearly resolved in the high-resolution STM scan of Figure~\ref{intermediates_summary}a (orange arrow)~\cite{briesenick2024kinetic}. Another notable feature is highlighted in  Figure~\ref{intermediates_summary}a by a dashed circle. It is a bright protrusion in the center of the TANG-TANG bond, which exhibits N-N internodal distances of 1.24$\pm$0.04~nm; this is longer than the measured TANG-TANG internodal distance in the rest of the structure (1.00$\pm$0.04~nm). We attribute this protrusion to an OM2 intermediate incorporated within the covalent polymer lattice on \ce{Au}(111), which are rarely observed~\cite{lischka2018surface,galeotti2019thiophene}. When imaged via constant-height AFM with a CO-decorated tip (Figure~\ref{intermediates_summary}b)~\cite{briesenick2024kinetic}, enhanced contrast of the molecular backbone is retrieved~\cite{Gross2009ChemicalStructureOfaMolecule}. However, the OM bridge atom that is readily seen in STM does not appear in AFM. We infer from the absence of contrast in the AFM images of the OM features that the molecules bend toward the Au surface (\textit{vide infra}). In contrast, the covalently bonded TANG-TANG linkages in the same region do not show such a protrusion in STM contrast, while constant-height AFM reveals a clear signature of the aryl groups in the frequency shift channel across the bonded region. 

\subsection{STM/AFM TANG Dimer Simulations}
In order to validate the interpretation of the contrast in the experimental STM/AFM images, VASP~\cite{Kresse1993Jan,Kresse1994May,Kresse1996Jul,Kresse1996Oct,Kresse1999Jan} DFT simulations (see Methods) of a partially brominated TANG dimer on \ce{Au}(111) were carried out. Figures~\ref{intermediates_summary}d--i summarize a series of dimer simulations using either an Au surface atom (OM1) or defect Au adatom (OM2). These represent the two possible OM coordination geometries.  Figure~\ref{intermediates_summary}d shows the OM2 intermediate with an in-plane N-N distance of 1.241~nm;  Figure~\ref{intermediates_summary}g models the OM1 intermediate and has a N-N distance of 1.210~nm. For the P$^2$TANG monolayer, the expected distance is 1.018~nm. Adsorption heights of the OM2 and OM1 dimers (N-Au) are 0.357 and 0.371~nm, respectively. In our simulations, the z coordinates of the four peripheral carbon atoms were fixed to mimic the constraint imposed by their incorporation into an extended, approximately planar polymer lattice. This could alter other anticipated relaxation effects, such as a lifting of the central nitrogen atom. Coordination of the dehalogenated carbon atom to the Au surface causes a considerable (0.194~nm) lifting of the surface atom. Similar strong interactions for rigid metal-coordinated dimers have been explored, leading to surface atom extraction on Cu(111)~\cite{ebeling2018adsorption}, Ag(111) and Au(111)~\cite{barton2017formation}.
The molecular bending was quantified from the N'--N--C angle (where N and N' are the nitrogen atoms of two adjacent TANG nodes and C is the dehalogenated carbon atom of the TANG, Figure~S26).
For the Au surface-atom coordination geometry, the molecule bends markedly toward the surface, with a maximum angle of 18.7~\degree.
In the Au-adatom coordination geometry, this bending is still present but reduced to 9.7~\degree.
The adatom is also lifted to a distance of 0.269~nm with respect to the surface, compared to the \ce{Au}(111) vertical interlayer distance of 0.235~nm. We conclude that the dashed-circle feature in Figure~\ref{intermediates_summary}a is best described by simulations involving a \ce{Au} surface adatom (Figure~\ref{intermediates_summary}e). Full density-based model (FDBM) non-contact AFM simulations are presented in Figures~\ref{intermediates_summary}f and~\ref{intermediates_summary}i~\cite{Ellner2019} (see Methods). In the case of the Au adatom, a slightly larger negative frequency shift along the OM bridge is again in good agreement with the experimental AFM images. This is expected because an Au adatom would lead to a more attractive interaction with the \ce{CO} at the tip than the Au surface atom (Figs.~S14 and S15). 
\subsection{Polymer Edge Terminations}

\begin{figure}
    \centering
    \includegraphics[width=\linewidth]{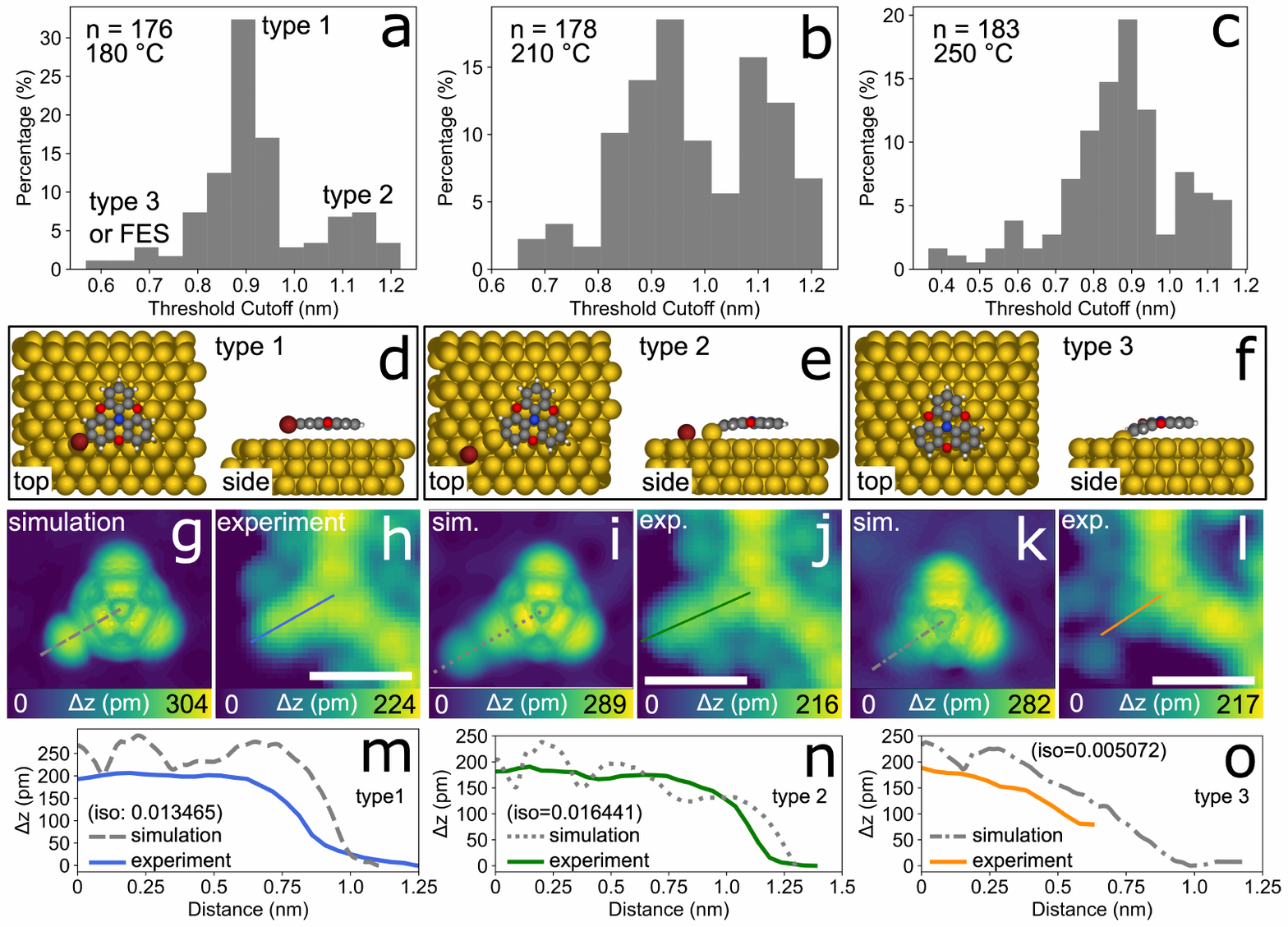}
    \caption{(a--c) Histograms showing the evolution of each termination grouping with varying $T_s$ ($T_c$ = 140~\si{\celsius}). (d--f) Simulation geometries for predominant P$^2$TANG edge unit terminated with a bromine (type 1), Au adatom (type 2) and Au surface atom (type 3) or featureless edge state. (g--l) Simulated constant-current STM alongside experimental examples. Experimental imaging parameters: $V_s$ = -100 mV, $I_t$ = 100 pA; simulation parameters: $V_s$ = -300 mV. Isosurface values: type 1, 0.013 e/\AA$^{3}$; type 2, 0.016 e/\AA$^{3}$; type 3, 0.005 e/\AA$^{3}$. Scale bars 1 nm. Growth parameters: $T_s$ = 210~\si{\celsius}, $T_c$ = 140~\si{\celsius}, $t_d$ = 3.5 min. (m--o) Experimental and simulated STM line profiles of each termination type.}
    \label{edge_terminations}
\end{figure}

We classified the polymer edge terminations into three different types based on their constant-current STM height decay profiles. This was done by extracting line profiles along polymer edges beginning at the central nitrogen atom and extending along the aryl group towards the Au surface as demonstrated by the solid line plotted in Figure~\ref{edge_terminations}h. The position at which the initial $z$ value reduced by 40\% was defined as the threshold cutoff. Histograms of these threshold cutoff values are plotted in Figures~\ref{edge_terminations}a--c for three sample anneal temperatures. For $T_s$ = 180~\si{\celsius}, the type 1 termination cutoff distance peaks at approximately 0.9~nm. Type 2 terminations (an example is shown in Figure~\ref{edge_terminations}j) are the second most prevalent and have a larger cutoff value ($\approx$ 1.15~nm). Type 3 terminations (Figure~\ref{edge_terminations}l) have a monotonically decaying profile moving away from the central nitrogen atom. They are considerably shorter ($\approx$ 0.7~nm) than either types 1 or 2. Both type 2 and 3 terminations exhibit lower apparent heights relative to the surrounding polymer, indicating molecular bending toward the Au surface.

To assign chemical identity to these termination types, we compare experimental line profiles with DFT-simulated STM for candidate edge geometries. The best matching geometries for each termination category are summarized in Figures~\ref{edge_terminations}d--f (see also Figure~S8 Supporting Information). Constant-current STM simulations for each type are plotted in Figures~\ref{edge_terminations}g,i,k and line profile comparisons are plotted in Figures~\ref{edge_terminations}m--o. For type 1, we find the best agreement for a bromine termination. Independent support comes from a similarly bright and rounded protrusion near the polymer termination in Figure~\ref{intermediates_summary}a (indicated by a white arrow), for which the line profiles are plotted in Figure~\ref{intermediates_summary}c. The AFM profile shows a peak $\approx$~0.65 nm from the central nitrogen atom. The spatial AFM contrast in Figure~\ref{intermediates_summary}b (white arrow) reveals an oval-shaped, bright lobe at the apex of the termination (of which several are visible in the whole scan frame). This contrast is consistent with a $\sigma$-hole signature of halogens as reported for related molecular systems on Au(111)~\cite{tschakert2020surface} and with AFM simulated images. Our calculations show that the C--Br bond places bromine approximately 0.320~nm above the average Au surface plane, compared to 0.190~nm for bromine adsorbed to a Au(111) hollow fcc site. This elevated position contributes to the bright contrast observed in STM at the polymer edges. As a related observation, this height difference also explains why bromine atoms within the pores of the film appear in STM contrast, but are absent in AFM constant-height images (Figure~S13). Additional constant-height STM data (Figure~S10) are also accordant with the termination in Figure~\ref{edge_terminations}h being bromine. Across 9 type 1 terminations examined with complementary AFM, all were consistent with bromine termination. 

This consistency has notable implications for type 2 terminations, which agree best with a polymer edge bonded to a Au adatom which is coordinated to an adjacent bromine atom (Figure~\ref{edge_terminations}n). Past studies found that the presence of adsorbed halogens is correlated to the release of adatoms from the Au(111) herringbone reconstruction which can then be incorporated into the resulting organometallic polymer structure~\cite{saywell2014manipulating, baker2010insights}. One study exploring the stabilization of organic tetracene molecules on Au(111) found that organometallic polymer-adatom structures are stabilized by surface-chemisorbed bromine attracted to the positively-charged Au adatoms~\cite{basagni2015surface}. 
The average STM height profile, which is measured from the center of the TANG molecule to the apex atom (bromine in type 1, Au adatom in type 2), is key to supporting the type 2 assignment: The type 2 profile is 0.02~nm lower than that of type 1, consistent with the molecule bending toward the surface in type 2, as predicated by DFT. This bending signature is also seen in constant-height STM (Figure~S10b). In STM, chains of adsorbed bromine atoms on the surrounding Au surface are frequently observed extending toward the apex (Figure~\ref{edge_terminations}j) of type 2 terminations, which is a proposed association with the type 2 termination, though direct verification by STM/AFM is not definitive regarding the species at the apex. This may suggest that the coordination of bromine may stabilize the polymer-Au adatom bond, which is common in homogeneous transition metal catalysis in solution~\cite{fagnou2002halide}. 

\begin{figure}
    \centering
    \includegraphics[width=0.5\linewidth]{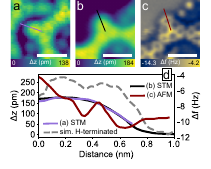}
    \caption{Examples of featureless edges not stabilized by surface atoms. (a) STM image from the same growth as the terminations plotted in Figure~\ref{edge_terminations} (-100 mV, 100 pA). (b) STM (-300 mV, 50 pA) image and (c) constant-height AFM ($f_0$ = 24,304 Hz, $A$ = 55 pm, $Q$ = 190,385) image showing the planar adsorption of such a termination. Scale bars 1 nm. (d) Line profiles taken from a--c.}
    \label{featureless_edge}
\end{figure}

Last, for type 3, the least commonly observed termination, the best agreement between experiment and simulation is obtained for a geometry in which the dehalogenated carbon atom is stabilized by a surface Au atom. This assignment is supported by the pronounced height decay in the simulated STM profile from the TANG core toward the edge, which is consistent with the strong downward bending toward the Au surface predicted by DFT. However, not all terminations classified by the cutoff analysis match the contrast in Figures~\ref{edge_terminations}k and \ref{edge_terminations}l and therefore are not type 3. We refer to these terminations as featureless edge states (FES). They appear to be bare polymer edges whose terminating species cannot be identified, as it is not resolved in either the STM or AFM images, as shown in Figure~\ref{featureless_edge}a–c~\cite{briesenick2024kinetic}. Figure~\ref{featureless_edge}d plots line profiles of an example FES. Passivation by residual hydrogen within the vacuum chamber is one possible assignment for the FES, though this is not directly verifiable from AFM contrast alone (Figure~S25). Hydrogen passivated edges could present a terminal blockage to further polymerization, but, as with type 3, they compose only a small number of the total termination population. For example, at 210~\si{\celsius}, of ten edge terminations with threshold cutoff values $\leq$0.75~nm, two were identified as type 3 and eight were determined to be FES. Future experiments involving deliberate \ce{H2} exposure and the population of FES could be used for further confirmation of a passivation mechanism. Additional examples of each termination type for different growth temperatures and tip conditions are summarized in Figure~S9.

Evolution of the population of each termination type for different surface temperatures can be rationalized as follows. At 180~\si{\celsius}, polymer edges are predominantly type 1, consistent with the gradual onset of dehalogenation, while type 2 terminations are relatively uncommon. As the temperature increases to 210~\si{\celsius}, the relative population of type 2 terminations increases. Polymer edges are now almost equally terminated by Au adatoms as well as unreacted bromine, and, simultaneously, the population of surface-diffusing bromine also increases. These measurements could be supplemented by real-time studies to capture the full reaction dynamics and influence of metal adatoms as demonstrated by Dettmann et al.~\cite{dettmann2024real}.
\subsection{Bromine Persistence}
\begin{figure}
    \centering
    \includegraphics[width=\linewidth]{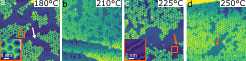}
    \caption{STM images (50 nm) showing partial coverage for the same $T_c$ = 140~\si{\celsius} and varying $T_s$. Insets in (a) and (c) are taken from the boxed regions. Chains of bromine atoms are indicated by orange arrows in (c) and (d). Imaging parameters: (a) -100 mV, 250 pA; (b) -300 mV, 20 pA; (c) -300 mV, 30 pA; (d) -300 mV, 30 pA.}
    \label{temperature_series}
\end{figure}

A set of $50\times50$~nm\textsuperscript{2} STM scans in Figure~\ref{temperature_series} show P\textsuperscript{2}TANG islands across the full range of explored anneal temperatures. Bromine, with a diffusion energy barrier of 0.09~eV on \ce{Au}(111)~\cite{andryushechkin2018adsorption}, is highly mobile, and should diffuse appreciably across all growth temperatures. This is evidenced by the nucleation of chains of bromine atoms which can be seen preferentially on the fcc sites of the \ce{Au} surface (Figure~S5). Despite the low diffusion barrier, the growth at 180~\si{\celsius}, shown in Figure~\ref{temperature_series}a, does not feature bromine on the surface outside the polymer pores or edges. One explanation is that the concentration of available bromine is below the threshold of nucleation into independent adlayers as most of the polymer edges remain halogenated (type 1). Halogen adlayers themselves have been extensively studied on \ce{Au}(111)~\cite{magnussen1996situ,huang1997determination,zheltov2014structural} and at sufficiently high surface coverage the Au surface reconstruction is heavily modified through substrate lattice deformations. Such a distortion is seen as a bending of the herringbone pattern around the polymer islands in Figure~\ref{temperature_series}a (white arrow).

Surface-bound bromine persists at 250~\si{\celsius} in our system, both as polymer edge terminations (type 1) and as chemisorbed adlayer structures (Figures~\ref{temperature_series}c and \ref{temperature_series}d, orange arrows) on the Au(111) surface. This is consistent with theoretical predictions that desorption from Au(111) occurs at appreciable rates above 525~\si{\celsius}~\cite{bjork2013mechanisms} and with quadrupole mass spectrometry measurements showing no associative \ce{Br2} desorption during on-surface polymerization~\cite{bronner2014tracking}. However, reported bromine desorption temperatures from temperature-programmed XPS studies vary.~\cite{mairena2019fate} Complete thermal desorption of bromine has been observed as low as $\approx$~225~\si{\celsius} for some precursors~\cite{fritton2019role} while others report onset to completion over the range of 300--400~\si{\celsius}~\cite{galeotti2019thiophene} or near-complete desorption at 250~\si{\celsius}~\cite{lischka2018surface}. This spread suggests that the removal of bromine is not governed solely by thermal desorption and the halogen-metal interaction. Total coverage, polymer confinement, interactions with the polymer edge sites, or residual \ce{H2} pressure in the vacuum chamber may play a role in bromine removal.
\subsection{Network Quality Quantification}
\begin{figure}
    \centering
    \includegraphics[width=\linewidth]{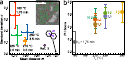}
    \caption{Quantitative comparison of network regularity using computational scoring methods based on MST and PH for P$^2$TANG networks for varying $T_s$. All growths used a $T_c$~=~140~\si{~\celsius}. (a) Normalized (by the square root of the cell area) mean segment length $\bar{m}$ vs. $\sigma$. The bars represent the spread over several scored images. The inset is an output of the MST scoring where the green colored zones are identified to have six nearest neighbors. The data point for Ref.~\cite{galeotti2020synthesis} is approximated from Figure~S14 of that work. (b) PH scores with $t_d$ given in minutes next to each data point.}
    \label{scoring_summary}
\end{figure}

Quantifying the order of each 2DCP is essential for comparing the quality under different deposition conditions. We assess the quality of P\textsuperscript{2}TANG films using STM constant-current scans and two established approaches to assess order/disorder. The first approach follows a minimum spanning tree (MST) approach~\cite{dussert1986minimal,wallet1998comparison}. Notably, this approach has already been used to assess the quality of 2DCPs including P\textsuperscript{2}TANG~\cite{ourdjini2011substrate,galeotti2019thiophene,galeotti2020synthesis}. The MST methodology relies on connecting a collection of points on a surface (image) with a continuous, branching line. This line is composed of segments of length $m$ which have an average value $\bar{m}$ and standard deviation $\sigma$. These two values then define the MST quantification. In our images, the points to be connected are the centers of the pores within the polymer network. A perfectly hexagonal P\textsuperscript{2}TANG network will have a normalized $\bar{m} \approx 1.07$ and $\sigma$ = 0. The results of the MST analysis are summarized in  Figure~\ref{scoring_summary}a. The $\bar{m}$ value of the film annealed at 250~\si{\celsius} is closer to that of an ideal network than the sample annealed at 180~\si{\celsius} which confirms the expected trend.

The second quantification method employs a persistent homology (PH) approach to assess image order~\cite{gutierrez2019persistent,obayashi2022persistent}. This topological analysis technique has been successfully applied to evaluate metal-organic frameworks under liquid and in vacuum conditions~\cite{lu2023order,cucinotta2025tuning}. The PH method identifies topological features in images and assigns these features so-called ``birth" and ``death" values. A feature's lifetime (death - birth), carries information about its shape and size. For each 2DCP, we quantified the polymer pores by plotting the lifetimes of all pores on a persistence diagram (birth vs. death). Pores with lifetimes falling within a defined similarity threshold constitute a subset of near-identical pores compared to the total feature population (Supporting Information Figure~S11). The final scalar PH score is calculated as the ratio of this subset to the total population. A maximum score of 1 would correspond to all pores being the same. Notably, this method is less sensitive to partial surface coverage and image boundaries, making it also robust for comparing samples with varying imaging conditions. The PH scores for films grown with a $T_c$ = 140~\si{\celsius} are summarized in Figure~\ref{scoring_summary}b. Notably, scored images for differing deposition times, and thus coverage, group together, highlighting robustness in the PH scoring method. PH scores for all growth parameter combinations are provided in Figure~S12. The PH scores indicate a steady improvement in network order as $T_s$ approaches 250~\si{\celsius}, while at lower temperatures we observe many more OM2 linkages (Supporting Information Figure~S4) and lower scores. 

\section{Conclusions}
We systematically studied the polymerization of TBTANG precursors into the polymer P\textsuperscript{2}TANG by means of on-surface synthesis on Au(111). We observed a previously unreported OM intermediate state involving Au adatoms coordinated within the growing polymer lattice. Such OM intermediates are frequently observed at the lowest annealing temperatures tested (180~\si{\celsius}) and proceed to decrease in number as the covalent network forms at elevated temperatures (250~\si{\celsius}). The disappearance of OM2 states is also accompanied by a quantitative increase in the final polymer order as determined from STM imaging paired with accepted MST and PH scoring techniques.

Additionally, we classified in detail the predominant polymer edge terminations as a function of sample anneal temperature. Using STM, high-resolution AFM, and DFT, we definitively identified the chemical structure of each termination. A majority of polymer edge states remain brominated up to 250~\si{\celsius}, corroborating a wide temperature range for the dehalogenation reaction as seen by other studies on Au(111). A large number of polymer edges are stabilized by Au adatoms which are accompanied by an adjacent bromine atom. This termination would seem to suggest the role bromine has in the polymer edge bonding to a Au adatom. Last, a small percentage of edges become stabilized to the Au surface itself, or present as featureless edges with no identifiable terminating species but are likely to be passivated by hydrogen.

This work emphasizes the necessity to explore the growth parameter space of precursor molecules in on-surface synthesis and specifically surface Ullmann coupling reactions. A wide temperature range for the removal of halogens and unknown coupling rates along with tunable deposition rates can lead to a highly variable final polymer quality. Predictability, repeatability, and quantification of polymer film quality by local probe methods are prerequisites for any future applications.

\section{Methods}
\subsection{Sample Preparation and STM/AFM Measurements}
Single-crystal \ce{Au}(111) surfaces were readied by means of \ce{Ar}$^+$-ion sputtering at energies of 0.7--1.0 kV and subsequent annealing cycles at 480 \degree C for 30 min. Molecular precursors (detailed synthesis procedures may be found in the Supporting Information) were sublimated from a commercial (Kentax) evaporator equipped with quartz crucibles in a dedicated UHV preparation chamber ($5 \times 10^{-10}$ mbar). Typical crucible temperatures ranged from 120--140~\degree C which corresponds to approximately 0.02--0.2 monolayers/min. Evaporator-sample distance and sample positions during deposition were kept consistent across all growths to maximize reproducibility. All molecular depositions were performed using a ``hot dosing" approach~\cite{galeotti2020synthesis}, where the clean \ce{Au}(111) surface was held at the desired polymerization temperature while depositing the precursors. The \ce{Au} substrate was held at the chosen growth temperature for 15 minutes after shuttering the deposition source, and cooled to room temperature over the course of 1 hour. The sample was then transferred directly to the measurement chamber without breaking vacuum and cooled to the base measurement temperature over the course of several hours before imaging. Experimental STM and non-contact AFM data were acquired with a commercial (Scienta Omicron) system. Measurements were performed under UHV with a base pressure of $1 \times 10^{-10}$ mbar and a base temperature of 9.6 K. Dual function STM/AFM sensors equipped with sharpened tungsten tips were employed to enable switching between imaging modes. Before measurement, tips were prepared on a clean Au surface through means of voltage pulses and controlled contact. High-resolution AFM images were acquired by exposing the sample to $1\times 10^{-9}$ mbar partial pressure of CO from a leak valve in the measurement chamber for 60 seconds. Functionalization of the gold-coated tungsten tip occurs spontaneously after dosing or scanning adsorbed CO in STM feedback. Experimental AFM images were denoised with a Gaussian filter.
\subsection{First-principles calculations}
All calculations were performed using Density Functional Theory as implemented in VASP (version 5.4.4)~\cite{Kresse1993Jan,Kresse1994May,Kresse1996Jul,Kresse1996Oct,Kresse1999Jan}.
The projector--augmented wave (PAW) method~\cite{Blochl1994Dec} was used to describe valence electrons of the atomic species: H $(1s)$,  C $(2s,2p)$, N $(2s,2p)$, O $(2s,2p)$, Br $(4s,4p)$ and Au $(5d,6s)$.
The generalized gradient approximation (GGA) was employed using the Perdew--Burke--Ernzerhof (PBE) exchange--correlation functional~\cite{Perdew1996Oct}.
Dispersion interactions corrections were included through the DFT-D3 approach~\cite{Grimme2010Apr,Grimme2011May}.
In selected cases, candidate adsorption geometries were pre-optimized using the MAD-SURF machine-learning interatomic potential~\cite{Lastre2026Jan_MADSURF} prior to the DFT calculations.

\subsection{AFM and STM image simulations}

Constant-height AFM images were simulated using the Full Density Based Model (FDBM)~\cite{Ellner2019, Zahl2021Nov, Ventura-Macias2023Oct} as implemented in the \texttt{DBSPM} GitHub repository:
\href{https://github.com/SPMTH/DBSPM}{\texttt{https://github.com/SPMTH/DBSPM}}.
The FDBM describes the probe--surface interaction as the sum of short-range Pauli repulsion, electrostatic interactions, and van der Waals dispersion,
\[
V_{\mathrm{total}} = V_{\mathrm{SR}} + V_{\mathrm{ES}} + V_{\mathrm{D3-vdW}},
\]
where the short-range and electrostatic contributions are evaluated from the DFT-calculated probe and surface charge densities and the surface electrostatic potential~\cite{Ellner2019}.
The parameters $\alpha$ and $V_0$ for the short-range Pauli-repulsion term were set to the universal values proposed in Ref.~\cite{Zahl2021Nov}, namely $\alpha = 1.08$ and $V_0 = 42.91~\mathrm{eV/\AA^{3(2\alpha-1)}}$.

Relaxation of the CO-functionalized probe was included using the default parameters implemented in \texttt{DBSPM}. At each tip position, the CO orientation was obtained by minimizing the probe--surface interaction together with a harmonic restoring potential that penalizes tilting of the CO away from its unperturbed orientation. A torsional stiffness of $\kappa = 0.20~\mathrm{eV\,rad^{-2}}$ and a probe pivot length of $3.02~\AA$ were used, with the probe orientation determined by Powell minimization.

The resulting force field was converted into frequency-shift images using an oscillation amplitude of $A = 0.55~\AA$ (55~pm), a free-oscillation resonance frequency of $f_0 = 30300~\mathrm{Hz}$, and a sensor stiffness of $k_0 = 1800~\mathrm{N\,m^{-1}}$.

STM images were simulated within the Tersoff--Hamann approximation~\cite{TersoffHamann}, where the tunneling current $I$ at a given bias voltage $V$ is proportional to the integrated local density of states (LDOS) of the sample at the tip position,
\[
I(\mathbf{r}, V) \propto \int_{E_F}^{E_F+eV} \rho(\mathbf{r}, E)\, dE ,
\]
with $E_F$ the Fermi energy and $\rho(\mathbf{r}, E)$ the LDOS of the sample at energy $E$.
In practice, this is obtained from the partial charge densities calculated by VASP in the corresponding energy window.
Constant-height images correspond to maps of this integrated density at a fixed tip--sample separation, while constant-current images are generated by adjusting the tip height such that the integrated density remains constant across the surface, reproducing the feedback loop of the STM experiment.

\begin{acknowledgement}

M.G.L., R.P. and P.P. acknowledge support from the Spanish Ministry of Science, Innovation and Universities, through project PID2023-149150OB-I00, from the  ``Mar\'{\i}a de Maeztu'' Programme for Units of Excellence in R\&D (CEX2023-001316-M) and from the Spanish Supercomputing Network (RES) for computational resources at the MareNostrum Supercomputer (BSC, Barcelona).

M.G.L. acknowledges support from the Spanish Ministry of Science, Innovation and Universities, through the predoctoral research contract PRE2021-098697. 

P.G. benefits from their RQMP membership https://doi.org/10.69777/309032. D.F.P. and P.G. acknowledge a FRQNT Team Grant. C.W.K. acknowledges support from the FRQNT Postdoctoral Fellowship. Funding from NSERC and CFI is gratefully acknowledged.

During the preparation of this work the authors W.A.B., M.G.L. and P.P. used ChatGPT (OpenAI) versions 4o and 5 and Claude (Anthropic, accessed via Duck.ai) version 4.6 for the purposes of improving text clarity and grammar. After using these services, the authors reviewed and edited the text as needed and take full responsibility for the content of the publication.

\end{acknowledgement}

\section*{Associated Content}

\subsection*{Data Availability Statement}
The experimental STM and AFM data and DFT calculation files
supporting this study are publicly available in Zenodo at
\url{https://zenodo.org/records/24395709}.
The repository includes experimental STM and AFM data, atomic
geometries, VASP input files, and the Br-mediated junction
relaxation trajectory, organized by figure in the main article
and Supporting Information.

\subsection*{Supporting Information}

Compound synthesis, purification, and HRMS characterization; additional STM data on rotational domains, monolayer coverage, defect densities, bromine adlayer formation, and edge terminations; details of the minimum spanning tree and persistent homology analyses used to quantify network quality; and additional AFM/STM simulations of bromine in polymer pores, organometallic intermediates, candidate edge-termination geometries, and molecular bending.

\bibliography{bibliography, UAM}

\end{document}


\tableofcontents
\clearpage

\section{Materials}
\subsection{Synthesis of 4,4',4'''-Tribromo-2,2':6',2'':6'',6-trioxatriphenylamine (TBTANG)}

\begin{figure}
    \centering
    \includegraphics[width=0.9\linewidth]{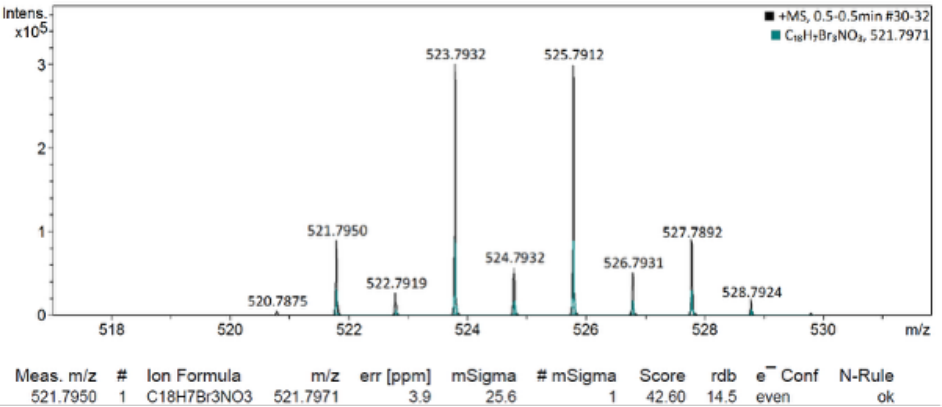}
    \caption{High-resolution mass spectrometry (HRMS) of TBTANG. APCI-MS (positive mode): calcd. for \ce{C18H7Br3NO3} [M+H]\textsuperscript{+} = 521.7976; found 521.7950.}
    \label{fig:HRMS_TBTANG}
\end{figure}

TANG~\cite{hamzehpoor2024azatriangulene} (0.30 g, 1.0 mmol) was dissolved in benzene (60 mL). Pyridinium tribromide (5.01 g, 15.7 mmol) in ethanol (40 mL) was added to the reaction. The mixture was degassed by \ce{N2} bubbling for 5 min and refluxed overnight. After cooling to room temperature, the yellow precipitate was retrieved by filtration and washed with ethanol (30 mL) three times to afford a yellow solid (0.35 g, 64\%). The crude powder (60 mg) was further purified by fractional sublimation: the first fraction (240~\si{\celsius}, 270 torr over 6 h) was mostly a dibrominated byproduct (10 mg). APCI-MS (positive mode): calcd. for \ce{C18H7Br2NO3} [M]\textsuperscript{+} = 442.8793; found 442.8779. The main fraction was obtained by continuing the sublimation at 280~\si{\celsius} (270 torr) for 3 h, affording pure TBTANG as yellow crystals (51 mg, 84\%). APCI-MS (positive mode): calcd. for \ce{C18H7Br3NO3} [M+H]\textsuperscript{+} = 521.7976; found 521.7950 (Figure \ref{fig:HRMS_TBTANG}).

\section{Supporting Data}
\begin{figure}
    \centering
    \includegraphics[width=0.4\linewidth]{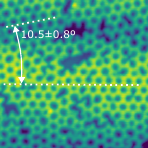}
    \caption{25 nm STM image ($V_s$ = -300 mV, $I_t$ = 50 pA) of P\textsuperscript{2}TANG showing two rotated domains growing together. Growth conditions: $T_s$ = 250~\si{\celsius}, $T_c$ = 140~\si{\celsius}, $t_d$ = 3.5 min.}
    \label{fig:two_domains}
\end{figure}

As reported previously from low-energy electron diffraction experiments~\cite{galeotti2020synthesis,dettmann2023electronic}, the P\textsuperscript{2}TANG polymer has two preferred rotational domains on \ce{Au}(111). The STM image in Figure~\ref{fig:two_domains} shows two polymer grains growing together with a rotational offset.

\begin{figure}
    \centering
    \includegraphics[width=0.7\linewidth]{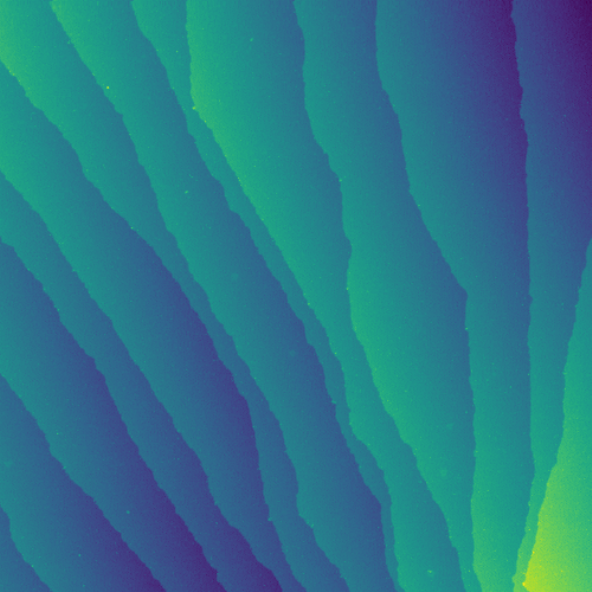}
    \caption{500 nm STM scan. Monolayer coverage achieved for the same deposition parameters as Figure~1b in the Main Text, but double the exposure time (3.5 min). Imaging parameters: $V_s$ = -300 mV, $I_t$ = 30 pA.}
    \label{fig:overgrowth}
\end{figure}

Figure~\ref{fig:overgrowth} shows monolayer (ML) coverage obtained for the same growth parameters as Figure~1b in the Main Text ($T_s$ = 180~\si{\celsius}, $T_c$ = 140~\si{\celsius}), but with twice the exposure time ($t_d$ = 3.5 min), which then corresponds to approximately 0.3 ML/min. We do not observe the nucleation or growth of a second layer.

\begin{figure}
    \centering
    \includegraphics[width=0.7\linewidth]{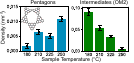}
    \caption{Densities of common defects within the P\textsuperscript{2}TANG network as a function of polymerization temperature ($T_c$ = 140~\si{\celsius}). Uncertainty estimates are based on Poisson counting statistics.}
    \label{fig:hand_count}
\end{figure}

To estimate the density of defects within the polymer network, we performed manual defect counting from STM images used in the MST scoring of the Main Text. Densities are weighted by the estimated polymer coverage as determined by the \textit{Gwyddion}~\cite{necas2012gwyddion} watershed masking process. Figure~\ref{fig:hand_count} reveals a larger number of pentagonal features at $T_s$ = 250~\si{\celsius} than at 180~\si{\celsius}. This is consistent with the formation of pentagons being a thermally limited process due to an increased strain energy cost. The number of OM intermediates within the network decreases as $T_s$ increases, following an increasing covalent coupling rate. 

\begin{figure}
    \centering
    \includegraphics[width=0.9\linewidth]{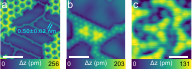}
    \caption{STM images of bromine adlayer features from a sample annealed at 250~\si{\celsius} (a) bromine chains forming between polymer regions. Scale bar 5~nm. (b) Two monomers presenting as bright triangular features trapped within the nucleating bromine. Scale bar 2~nm. (c) A TANG monomer trapped in a congested region of bromine atoms and surrounding polymer ($T_s$ = 180~\si{\celsius}). Scale bar 1~nm. Imaging parameters: (a) -300~mV, 30~pA; (b) 50~mV, 50~pA; (c) -100~mV, 250~pA.}
    \label{fig:halogen_features}
\end{figure}

In addition to bromine trapped within the pores of the growing polymer, we observe net-like features representing the formation of bromine adlayers on the Au(111) surface. Some examples are shown in Figure~\ref{fig:halogen_features}a. Rare instances of what we interpret to be monomers trapped within this growing network of bromine were also observed (Figures~\ref{fig:halogen_features}b,c). The presence of bromine adlayer structures confirms its persistence up to 250~\si{\celsius}.

\subsection{Edge Termination Classification}
\begin{figure}
    \centering
    \includegraphics[width=\linewidth]{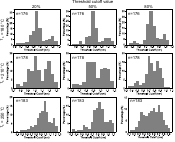}
    \caption{Histograms of the threshold values of the data presented in Figs. 3a-c of the Main Text with cutoff percentages of 20, 50, and 80~\%.}
    \label{fig:edge_termination_thresholds}
\end{figure}

\begin{figure}
    \centering
    \includegraphics[width=\linewidth]{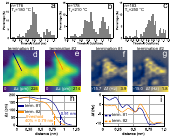}
    \caption{(a-c) Histograms using the same data as Figures 3a-c in the Main Text but with bin widths of 25~pm. (d,e) 2~nm STM images (-300~mV, 50~pA) of two terminations classified as type 1. (f,g) Corresponding AFM images ($f_0$ = 24,304 Hz, $A$ = 55 pm, $Q$ = 190,385) of the same terminations. (h) STM line profiles extracted from the lines plotted in (d,e). The 40~\% threshold cutoff value determined from the STM profile is 0.91~nm (0.79~nm) for termination \#1 (\#2). (i) AFM line profiles extracted from (f,g) starting at the central N atom extending outward towards the termination in the same fashion as (d,e).}
    \label{fig:edges_25pm_bins}
\end{figure}

To classify the polymer edge terminations, we extract line profiles from constant-current STM images and mark the threshold value where the $\Delta z$ value drops by a given percentage. A threshold value of 40\% is chosen for Figs. 3a-c in the Main Text; alternate percentage cutoff values are plotted in Figure~\ref{fig:edge_termination_thresholds}. The relative peak intensities are consistent for the 180~\ce{\celsius} and 210~\si{\celsius} data sets across various cutoff values and deviates for the 250~\si{\celsius} data set meaningfully at 80~\%. This can be explained by a difference in STM image resolution due to tip conditioning for the 250~\si{\celsius} data. Broadening and asymmetry of the type 1 peak is examined by reducing the bin size from 50~pm to 25~pm (Figures~\ref{fig:edges_25pm_bins}a-c). At 210~\si{\celsius} (Figure~\ref{fig:edges_25pm_bins}b) two peaks appear to emerge centered around $\approx$ 0.85~nm and $\approx$ 0.94~nm. STM images for two terminations from a $T_s$ = 210~\si{\celsius} growth are plotted in Figures~\ref{fig:edges_25pm_bins}d,e with edge termination line profiles plotted in Figure~\ref{fig:edges_25pm_bins}h. The 40~\% threshold cutoff values for termination 1 and 2 are 0.91~nm and 0.79~nm, respectively. These values fall within the type 1 peak. The respective constant-height AFM images for both terminations are shown in Figures~\ref{fig:edges_25pm_bins}f,g. The asymmetric bright protrusion is consistent with the type 1 (bromine) assignment for both edges. AFM profiles extracted in the same manner as the STM line profiles are plotted in Figure~\ref{fig:edges_25pm_bins}i. Variations in the exact position of the bromine over the underlying Au(111) lattice would account for changes in the STM contrast and thus changes in the cutoff threshold. Furthermore, broadening of the peaks may be assigned to uncertainties in the extraction of the line profiles' starting position, direction, or asymmetry in the fast/slow scan directions.
\subsection{Alternate Edge Terminations}
\begin{figure}
    \centering
    \includegraphics[width=0.7\linewidth]{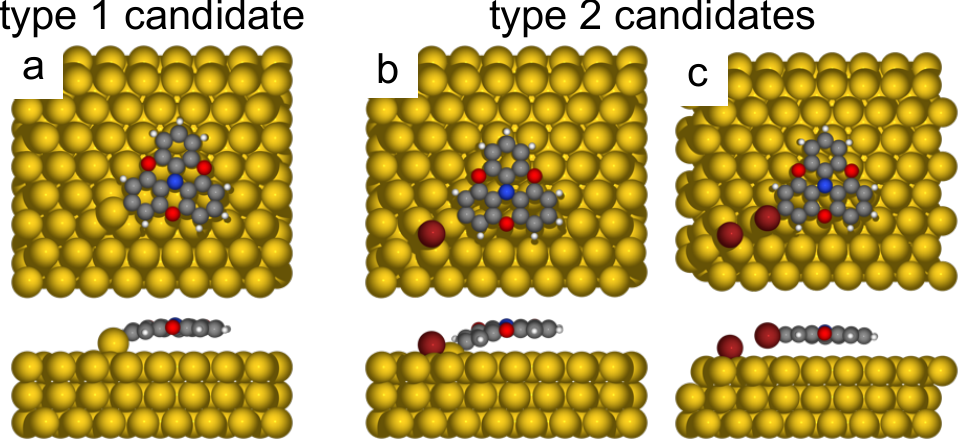}
    \caption{Simulated alternate candidate geometries for polymer edge terminations. (a) Type 1: a polymer edge stabilized by an Au adatom. (b) Type 2: stabilized to an Au surface atom with a chemisorbed bromine nearby (``Au surface, Br''). (c) Type 2: bromine terminated with an additional bromine near the apex (``C-Br, Br'').}
    \label{fig:alt_term_geometries}
\end{figure}

\begin{figure}
    \centering
    \includegraphics[width=\linewidth]{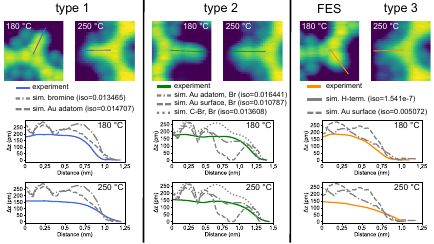}
    \caption{Additional experimental STM images (2~nm) of type 1 (left column), type 2 (center column), and type 3 or FES (right column) terminations for different $T_s$. Constant-current STM line profiles are plotted alongside all simulated edge terminations. Imaging parameters: 180~\si{\celsius}, CO tip, ($V_s$ = -100 mV, $I_t$ = 250 pA); 250~\si{\celsius}, metal tip ($V_s$ = -300 mV, $I_t$ = 30 pA). Isovalues: type 1, 0.013 e/\AA$^{3}$; type 2, 0.016 e/\AA$^{3}$; type 3, 0.005 e/\AA$^{3}$.}
    \label{fig:addition_terms}
\end{figure}

\begin{figure}
    \centering
    \includegraphics[width=\linewidth]{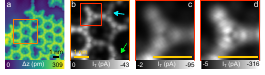}
    \caption{(a) Constant-current STM image ($V_s$ = -100 mV, $I_t$ = 100 pA) with a CO tip of a P$^2$TANG polymer segment from which the three termination types in Figure~3 were identified. (b) Constant-height STM image of the boxed region in (a) ($\Delta z = 200$ pm relative to a $-100$ mV, 50 pA setpoint). The type 1 termination shown in Figure~3h is outlined. Solid (dashed) arrows indicate representative type 2 (type 3) terminations. (c) Constant-height STM image of the boxed region in (b) at $\Delta z = 0$ pm. (d) Constant-height STM image at $\Delta z = -150$ pm.}
    \label{fig:CH_STM_data}
\end{figure}

Alternate polymer edge termination structures were modeled, and a summary of these is shown in Figure~\ref{fig:alt_term_geometries}. For type 1 terminations, one alternate geometry was modeled, which is that of a single Au adatom (Figure~\ref{fig:alt_term_geometries}a, STM/AFM simulations Fig.~\ref{fig:tbtang_terminated_Au_adatom_pp_dbspm_stm}). The left column of Figure~\ref{fig:addition_terms} compares simulated STM contrast from the considered geometries (as line profiles) for type 1 terminations to experimental examples from growths at $T_s$ = 180 and 250~\si{\celsius}. While the gradual decay profile of the Au adatom termination at distances $>$~0.6~nm does seem to match the experimental line profiles more closely, we indicated in the Main Text that the debromination is gradual at 180~\si{\celsius}, making bromine termination much more likely. Moreover, the STM image contrast is very similar to Figure~3h. Additionally, constant-height STM imaging of the same type 1 termination presented in Figure~3h is shown in Figures~\ref{fig:CH_STM_data}b--d, which shows minimal molecular bending and thus making an Au adatom bond very unlikely. The experimental constant-height STM is in agreement with the simulated STM contrast of Figure~\ref{fig:dbspm_stm_tbtang_br}. For the longer type 2 terminations, we considered two alternate geometries; these are a bonded Au surface atom with a chemisorbed bromine nearby (Figure~\ref{fig:alt_term_geometries}b, STM/AFM simulations Fig.~\ref{fig:tbtang_br_au_surface_dbspm_stm}) and a C--Br bond with a bromine nearby (Figure~\ref{fig:alt_term_geometries}c, STM/AFM simulations Fig.~\ref{fig:tbtang_br_br_surface_dbspm_stm}). Experimental and simulated STM comparisons for all considered type 2 geometries are plotted in the center column of Figure~\ref{fig:addition_terms}. We found poor agreement with the polymer edge bonded to an Au surface atom. Comparing the two remaining cases, from the line profiles alone it is difficult to assign the correct termination. Therefore, we again consider molecular bending at the film edge. In constant-current experimental and simulated images, we observe more bending towards the substrate surface in the case of the Au adatom even with an adjacent bromine nearby. This supports the conclusion that type 2 terminations consist of the polymer edge bonded to an Au adatom with an adjacent bromine nearby. Type 3 and FES terminations are plotted in the rightmost column of Figure~\ref{fig:addition_terms}.

\section{Quantification of Network Quality}
\subsection{Processing for Minimum Spanning Tree (MST) Analysis}
All STM images used in the MST analysis were identically preprocessed to increase comparability across polymer growths. To that end, only sample regions without \ce{Au} step edges were selected to avoid systematic errors. Raw STM images were plane-leveled in \textit{Gwyddion} to account for sample tilt~\cite{necas2012gwyddion}. The MST is then calculated using the code found here: 
\newline 
\href{https://github.com/lvbesteiro/STM-Minimum\_Spanning\_Tree}{\texttt{https://github.com/lvbesteiro/STM-Minimum\_Spanning\_Tree}}~\cite{galeotti2019thiophene,galeotti2020synthesis}. This code equalizes the grayscale image’s contrast using the \textit{skimage} package. Sobel edge detection is used to locate the molecular backbone structure. A watershed process fills and labels the polymer pores (foreground) to distinguish them from the background. The center of mass is then determined for each pore which is recorded in terms of pixel coordinates. The MST function from the \textit{NetworkX} package (\href{https://networkx.org/documentation}{\texttt{https://networkx.org/documentation}})
then operates on this collection of nodes, connecting the pores’ centers of mass using length segments that minimize the total path length between the nodes. The output of the MST algorithm is a collection of internodal segments of length $m$. The average length ($\bar{m}$) of these MST branches is calculated by adding the total length of all segments divided by the total number of segments. The standard deviation $\sigma$ is calculated in a similar fashion. Both $\bar{m}$ and $\sigma$ are dimensionless (normalized by the square root of the cell area) and contain information regarding the order of the 2D framework; a perfectly regular network will have a singular $\bar{m}$ with $\sigma = 0$~\cite{ourdjini2011substrate}. Inter-image uncertainties are estimated from a minimum of 3 images per growth condition.
\subsection{Processing for Persistent Homology (PH) Scoring}
\begin{figure}
    \centering
    \includegraphics[width=0.6\linewidth]{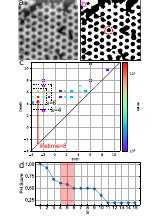}
    \caption{(a) Leveled STM image (15 nm, 150 px). (b) Otsu threshold binarized image. The red circle highlights an example of a pore with the most prevalent lifetime (birth = -3, death = 5, lifetime = 8). (c) Persistence diagram generated for the binarized image. The color scale represents the number of counts with a particular (birth, death) pairing. The vertical distance of a data point from the line of slope=1 represents that point's lifetime. The boxed areas highlights the search regions for a given $s$ value. (d) Scalar PH score computed from the persistence diagram in (c) as a function of $s$. The final PH score for the image is averaged from $s$ = 4--6 (highlighted values).}
    \label{fig:persistence_diagram}
\end{figure}

The order/disorder of polymer films is also quantitatively assessed using a persistent homology approach. The scoring process, outlined in Figure~\ref{fig:persistence_diagram}, involves generating a persistence diagram from STM images. First, leveled STM images (Figure~\ref{fig:persistence_diagram}a) are binarized by Otsu thresholding (Figure~\ref{fig:persistence_diagram}b). The persistence diagram (Figure~\ref{fig:persistence_diagram}c) of the binarized image is then computed using the \textit{HomCloud}~\cite{obayashi2022persistent} version 4.8.0 analysis software. This program’s filtration process identifies the pores in the processed STM data by groups of pixels, identified as the polymer backbone (shown here in white), that form closed loops around pixels identified as belonging to the background (black)~\cite{buchet2018persistent, gutierrez2019persistent}. The pores are assigned a birth time (when the feature first appears during filtering) and a death time (when it terminates or merges with another feature). A feature's lifetime corresponds to the difference between these two values. The lifetime contains information of the size and shape of each pore. All birth and death pairs are plotted (x vs. y) as a persistence diagram (PD), where all pairs exist above a line of x=y since a feature cannot “die” before its birth.

\begin{figure}
    \centering
    \includegraphics[width=0.7\linewidth]{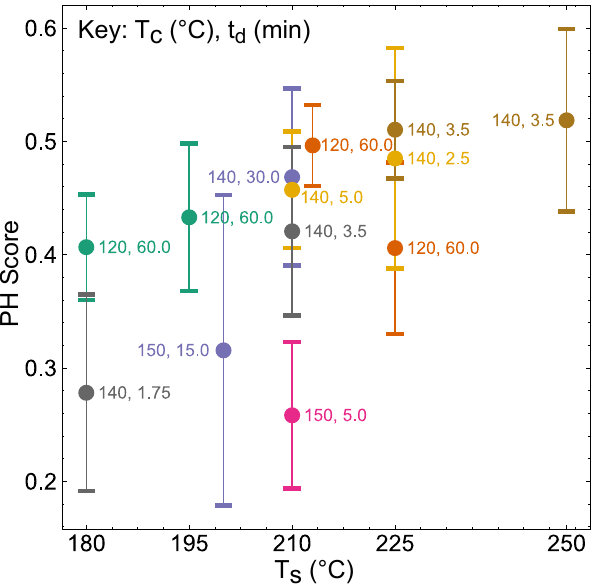}
    \caption{Computed PH scores for all P\textsuperscript{2}TANG growths.}
    \label{fig:all_PH_scores}
\end{figure}

Scalar PH scores are calculated by grouping features with similar lifetimes within a given PD. For a given search parameter $s$, each unique lifetime value has its neighboring features counted within $\pm$~lifetime~$\times~1/s$. For example, $s=4$ corresponds to a 25 \% similarity threshold. The number of features within this subset is then divided by the total number of identified features in the image to yield the PH score for that $s$ value. The overall score for the image (for a particular $s$ value) is given by the subset having the highest score. The computed scalar PH scores for the image in Figure~\ref{fig:persistence_diagram}a from $s=1\textrm{--}15$ are plotted in Figure~\ref{fig:persistence_diagram}d. PH scores from $s=4$ to $s=6$ are averaged to obtain a final score (highlighted region of Figure~\ref{fig:persistence_diagram}d). A higher final PH score implies that more of the pores in the image are alike, thus the structure in the image is more ordered. The standard deviation of the PH score across a sampling of several STM scans for each set of growth parameters provides the score’s uncertainty. Image pixel densities remained constant for scored images. PH scores for all film growths are shown in Figure~\ref{fig:all_PH_scores}. The trend towards higher order (higher PH score) is consistent with increasing $T_s$.

\clearpage

\section{Additional HR-AFM and STM simulations}
\begin{figure}
    \centering
    \includegraphics[width=1\linewidth]{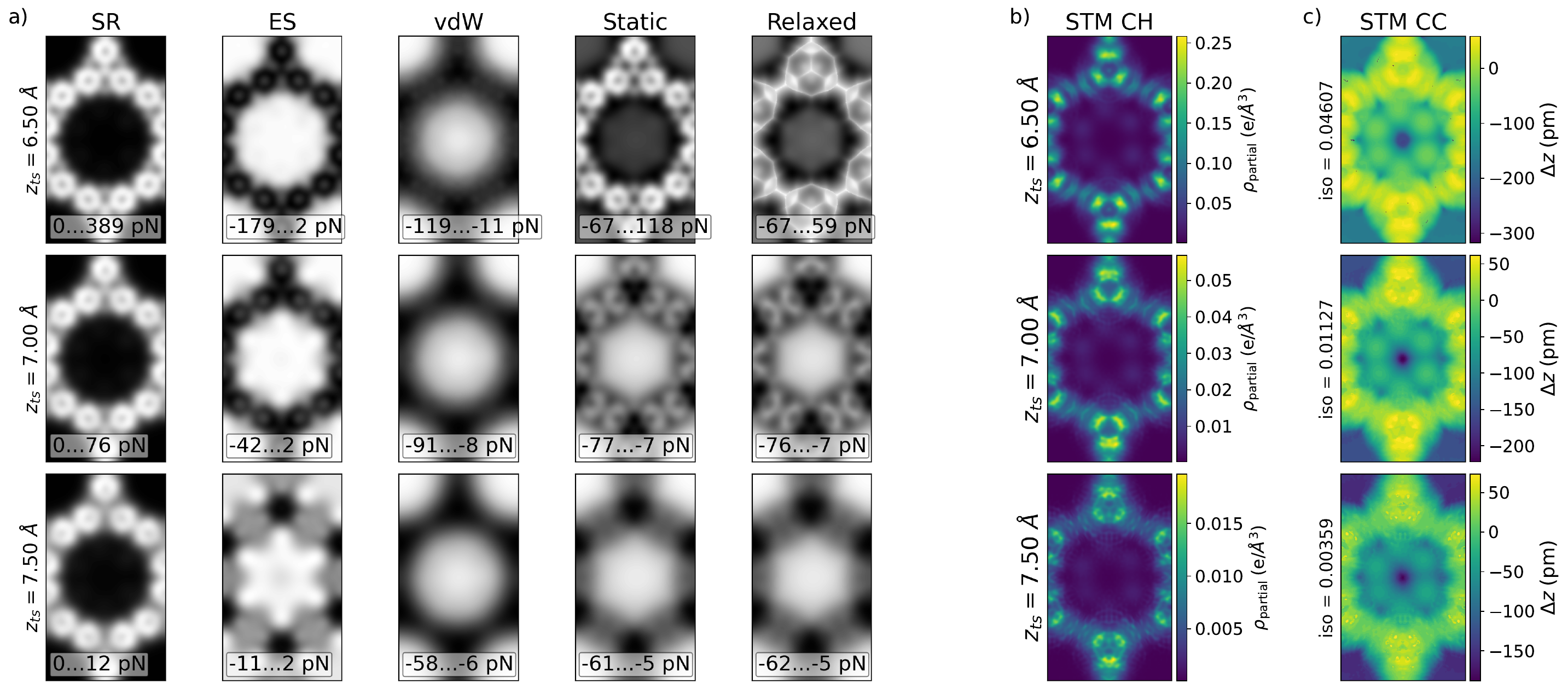}
    \caption{Combined HR-AFM and STM simulations for a rectangular P$^2$TANG supercell containing six adsorbed Br atoms trapped within a pore (cf. Fig.~1). (a) Full density-based AFM (FDBM) simulations decomposed into the short-range (SR), electrostatic (ES), van der Waals (vdW), static, and relaxed contributions at several tip heights relative to the Au(111) surface. (b,c) Corresponding constant-height and constant-current STM simulations. At the imaging heights considered, the HR-AFM contrast is dominated by the monolayer and does not resolve the adsorbed bromine atoms, whereas they are clearly visible in the STM signal.}
    \label{fig:monolayer_w_6_Br}
\end{figure}

\begin{figure}
    \centering
    \includegraphics[width=1\linewidth]{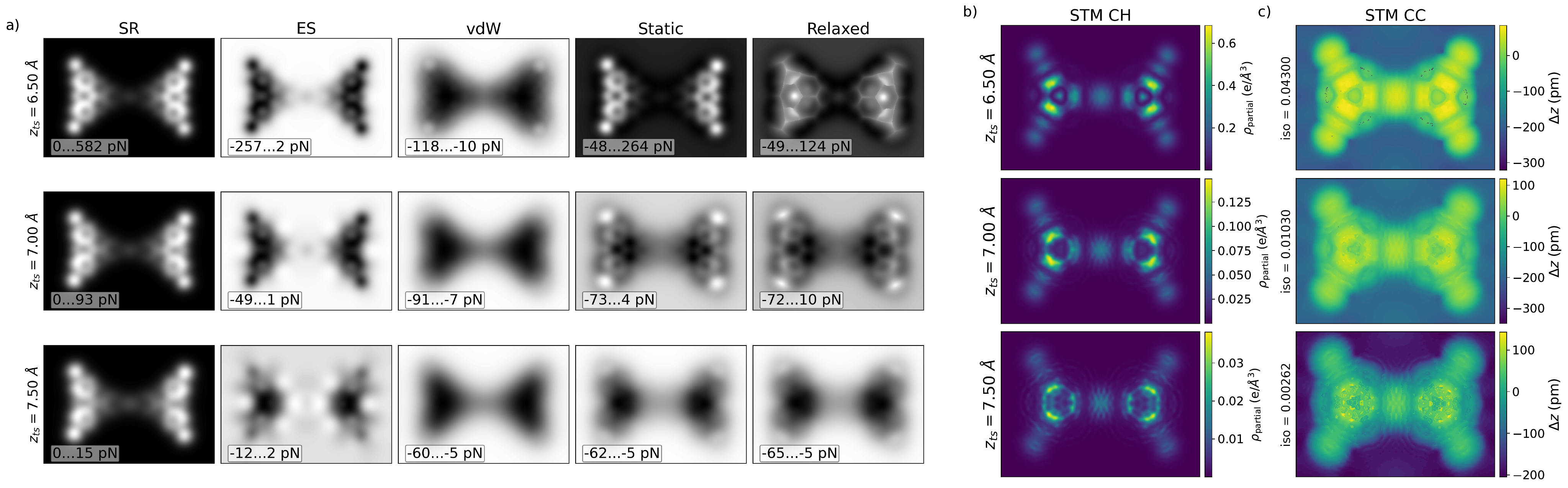}
    \caption{Combined HR-AFM and STM simulations for a partially brominated TANG dimer
coordinated by an Au adatom at a top site, corresponding to the OM geometry discussed in Fig.~2d. (a) Full density-based AFM (FDBM) simulations showing the short-range (SR), electrostatic (ES), van der Waals (vdW), static, and relaxed contributions at several tip heights with respect to the Au(111) surface. (b, c) Corresponding constant-height and constant-current STM simulations. This configuration reproduces the experimentally observed OM contrast most closely, particularly the enhanced signal at the bridge region.}
    \label{fig:tang_dimer_Au_adatom_top}
\end{figure}

\begin{figure}
    \centering
    \includegraphics[width=1\linewidth]{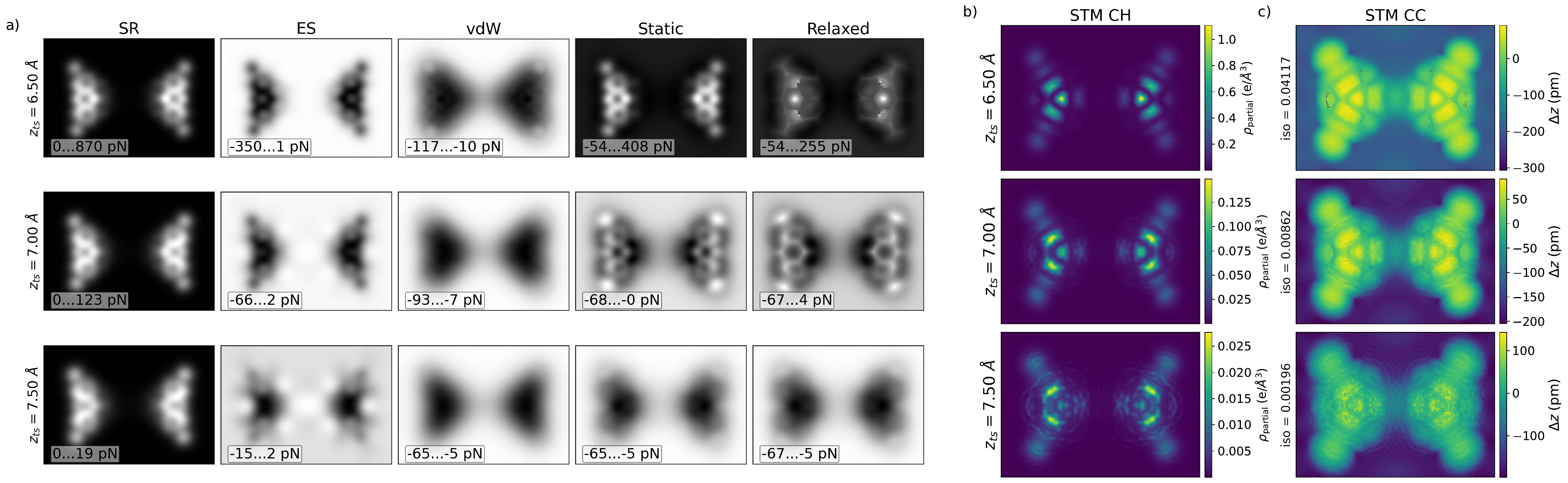}
    \caption{Combined HR-AFM and STM simulations for a partially brominated TANG dimer coordinated through an Au surface atom, corresponding to the alternative OM geometry discussed in Fig.~2g. (a) Full density-based AFM (FDBM) simulations showing the short-range (SR), electrostatic (ES), van der Waals (vdW), static, and relaxed contributions at several tip heights with respect to the Au(111) surface. (b, c) Corresponding constant-height and constant-current STM simulations. Compared with the Au-adatom model, this geometry yields a weaker bridge contrast.}
    \label{fig:tang_dimer_Au_surface_atom}
\end{figure}

\begin{figure}
    \centering
    \includegraphics[width=1\linewidth]{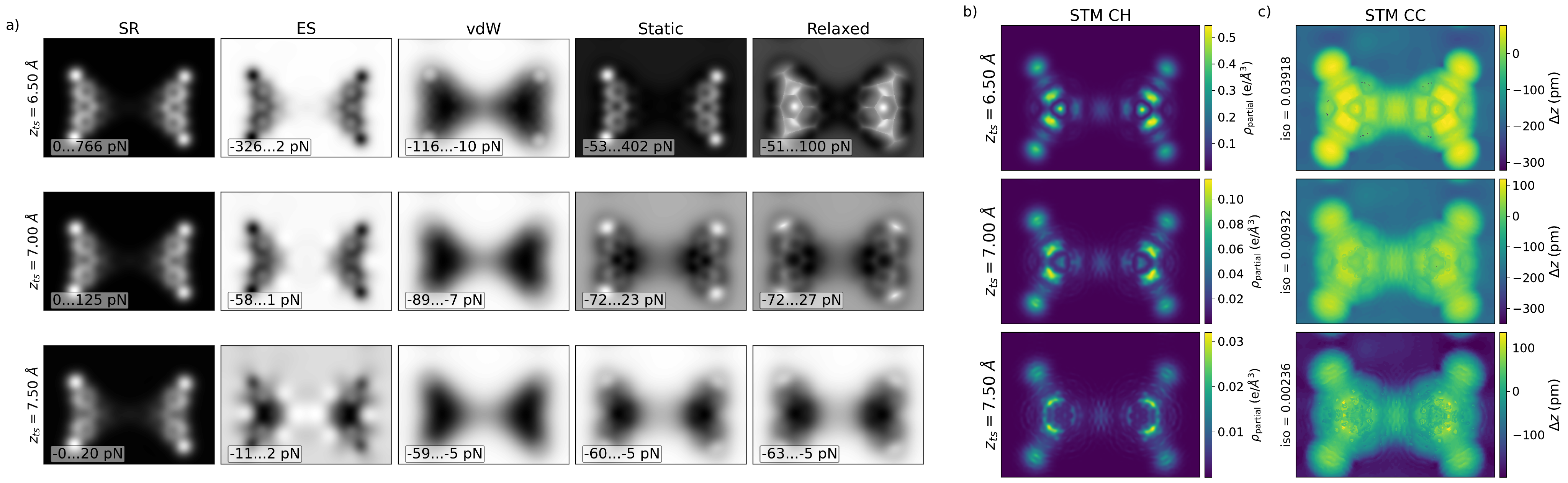}
    \caption{Combined HR-AFM and STM simulations for a partially brominated TANG dimer coordinated by an Au adatom in a hollow site. (a) Full density-based AFM (FDBM) simulations showing the short-range (SR), electrostatic (ES), van der Waals (vdW), static, and relaxed contributions at several tip heights with respect to the Au(111) surface. (b, c) Corresponding constant-height and constant-current STM simulations. This geometry produces an intermediate OM-junction contrast between the top-site Au-adatom model and the Au-surface-atom model.}
    \label{fig:tang_dimer_Au_adatom_hollow}
\end{figure}

\begin{figure}
    \centering
    \includegraphics[width=1\linewidth]{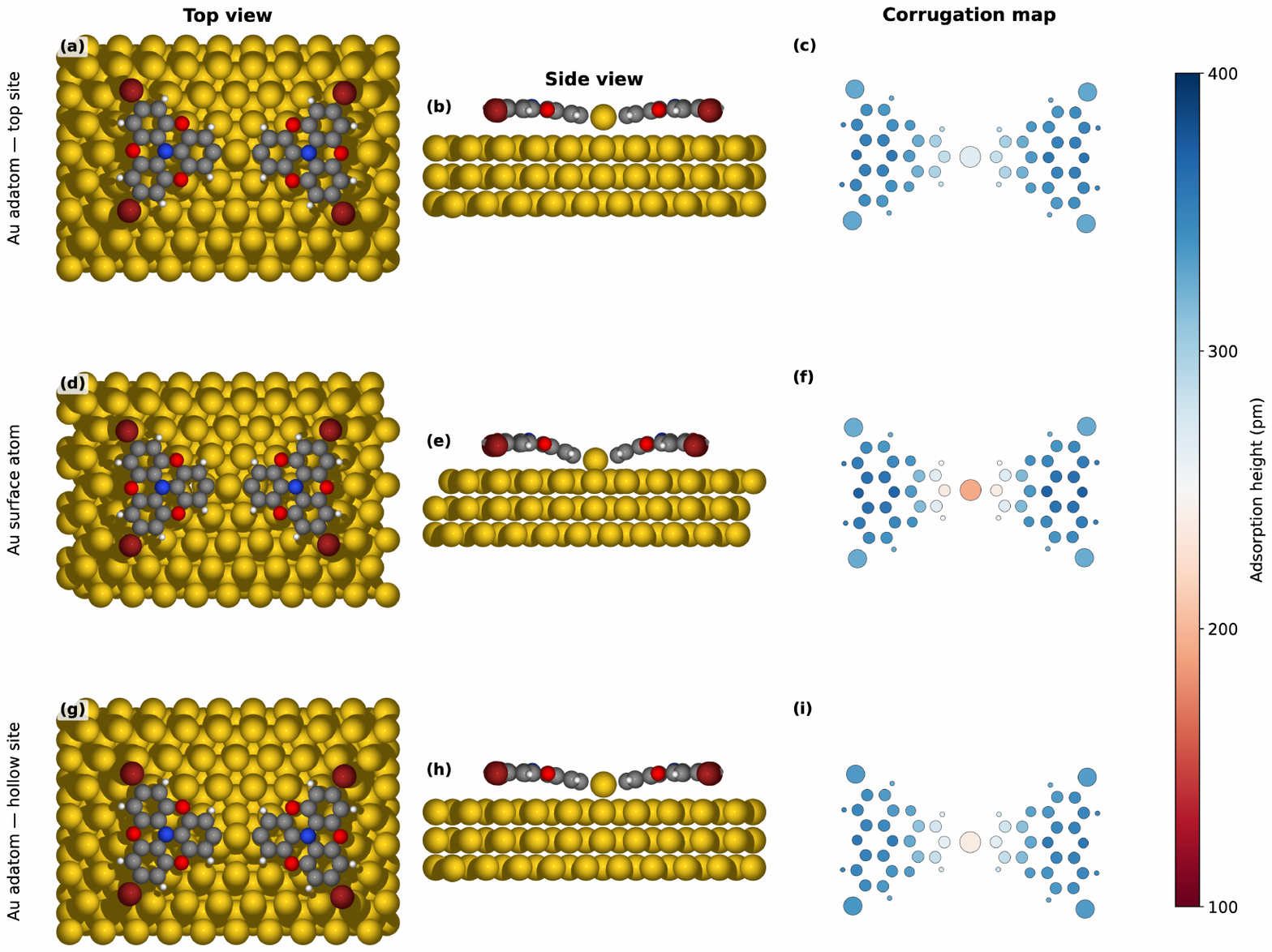}
    \caption{Comparison of the relaxed geometries and molecular corrugation of the three Au-mediated TANG dimer configurations considered for the organometallic junction in Figure~2 of the Main Text. (a--c) Au-adatom coordination with the adatom at a top site (OM2); (d--f) coordination through an Au surface atom (OM1); and (g--i) Au-adatom coordination with the adatom at a hollow site (alternative OM2 configuration). For each configuration, the left and center panels show top and side views, respectively, of the same relaxed geometry, while the right panel shows the corresponding corrugation map. The color scale represents the vertical adsorption height relative to the Au(111) surface plane. The Au-surface-atom geometry exhibits the strongest downward bending of the TANG units, whereas both Au-adatom configurations retain a more planar geometry. These structural differences are consistent with the molecular bending quantified in Figure~\ref{fig:angle_clarification} and help rationalize the differences between the simulated AFM contrasts of the candidate organometallic junctions.}
    \label{fig:OM_types_corrugation_maps}
\end{figure}

\begin{figure}
    \centering
    \includegraphics[width=1\linewidth]{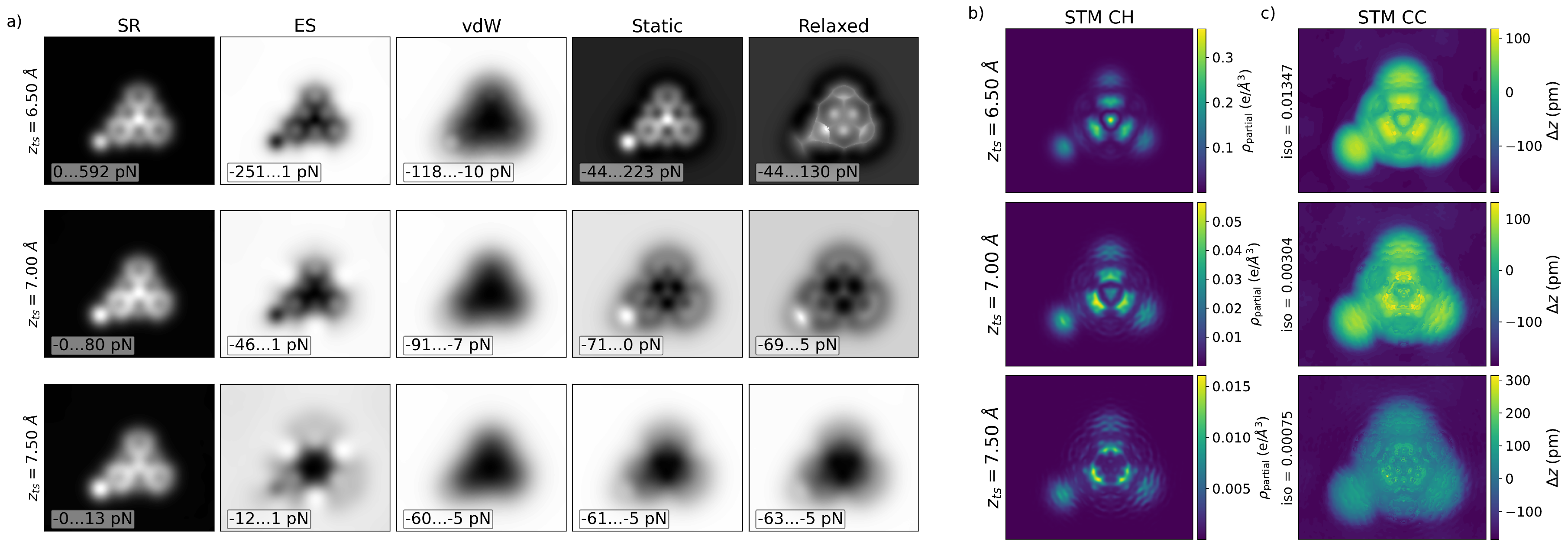}
    \caption{Combined HR-AFM and STM simulations for the bromine-terminated P$^2$TANG edge motif assigned to type 1 in Figure~3. (a) Full density-based AFM (FDBM) simulations showing the short-range (SR), electrostatic (ES), van der Waals (vdW), static, and relaxed contributions at several tip heights with respect to the Au(111) surface. (b, c) Corresponding constant-height and constant-current STM simulations.}
    \label{fig:dbspm_stm_tbtang_br}
\end{figure}

\begin{figure}
    \centering
    \includegraphics[width=1\linewidth]{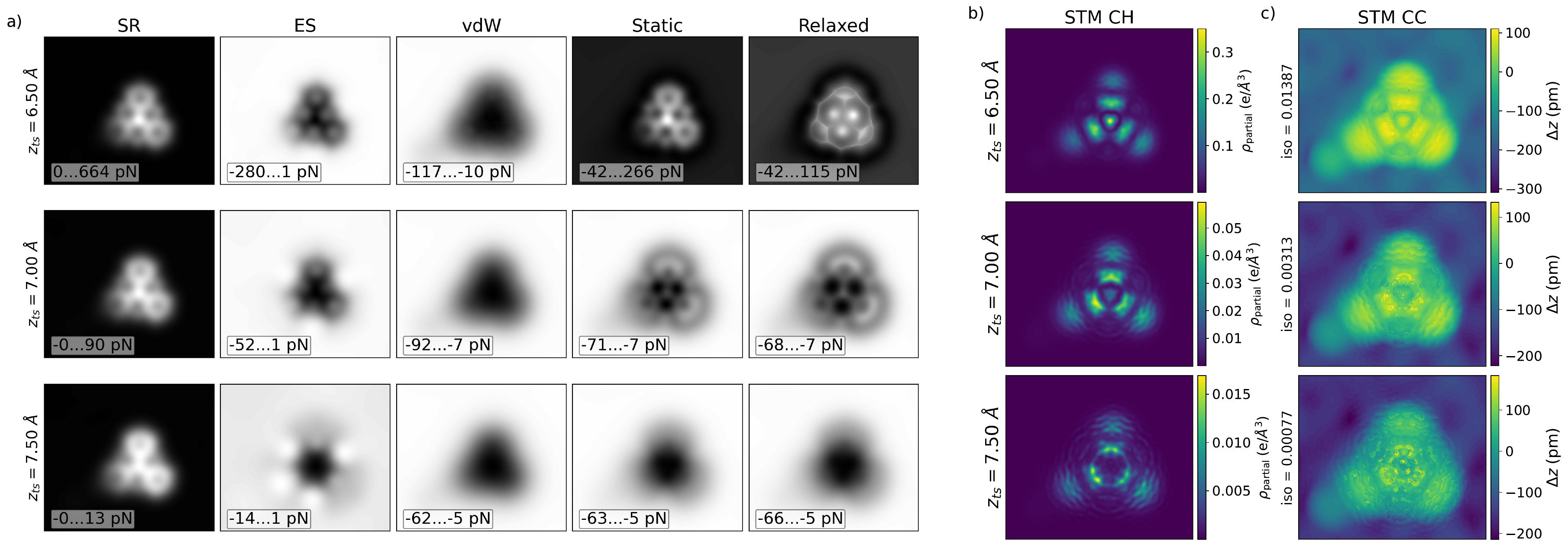}
    \caption{Combined HR-AFM and STM simulations for the P$^2$TANG edge unit stabilized by a gold adatom with a nearby bromine atom, assigned to type 2 in Figure~3. (a) Full density-based AFM (FDBM) simulations showing the short-range (SR), electrostatic (ES), van der Waals (vdW), static, and relaxed contributions at several tip heights with respect to the Au(111) surface. (b, c) Corresponding constant-height and constant-current STM simulations.}
    \label{fig:dbspm_stm_tbtang_br_au}
\end{figure}

\begin{figure}
    \centering
    \includegraphics[width=1\linewidth]{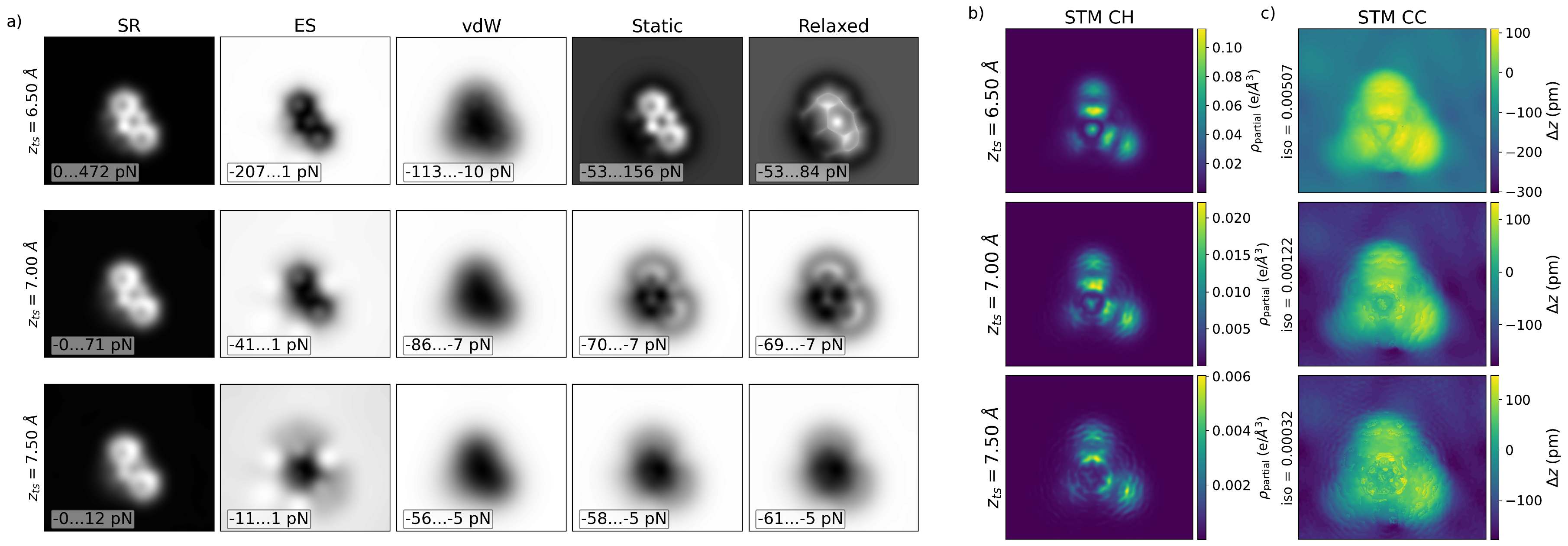}
    \caption{Combined HR-AFM and STM simulations for the P$^2$TANG edge unit stabilized by a gold surface atom, assigned to type 3 in Figure~3. (a) Full density-based AFM (FDBM) simulations showing the short-range (SR), electrostatic (ES), van der Waals (vdW), static, and relaxed contributions at several tip heights with respect to the Au(111) surface. (b, c) Corresponding constant-height and constant-current STM simulations.}
    \label{fig:tbtang_aryl_pp_dbspm_stm}
\end{figure}

\begin{figure}
    \centering
    \includegraphics[width=1\linewidth]{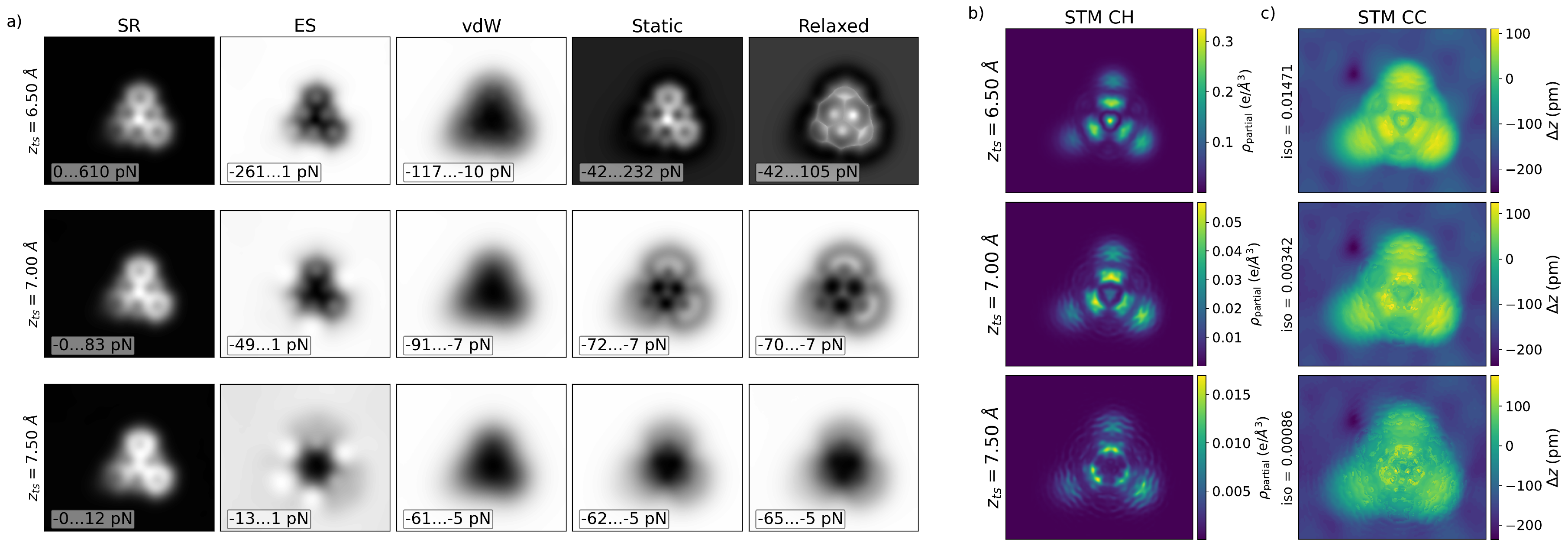}
    \caption{Combined HR-AFM and STM simulations for the P$^2$TANG edge unit stabilized by a gold adatom (Fig.~\ref{fig:alt_term_geometries}a). (a) Full density-based AFM (FDBM) simulations showing the short-range (SR), electrostatic (ES), van der Waals (vdW), static, and relaxed contributions at several tip heights with respect to the Au(111) surface. (b, c) Corresponding constant-height and constant-current STM simulations.}
    \label{fig:tbtang_terminated_Au_adatom_pp_dbspm_stm}
\end{figure}

\begin{figure}
    \centering
    \includegraphics[width=1\linewidth]{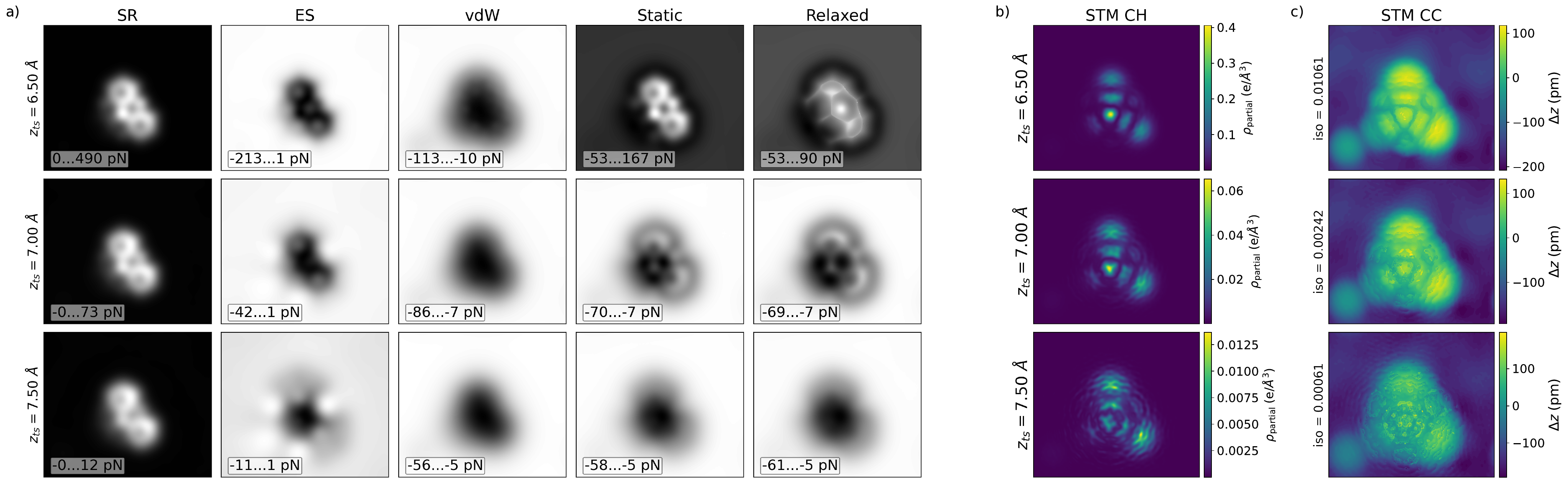}
    \caption{Combined HR-AFM and STM simulations for the P$^2$TANG edge unit stabilized by a gold surface atom with a Br atom nearby (Fig.~\ref{fig:alt_term_geometries}b). (a) Full density-based AFM (FDBM) simulations showing the short-range (SR), electrostatic (ES), van der Waals (vdW), static, and relaxed contributions at several tip heights with respect to the Au(111) surface. (b, c) Corresponding constant-height and constant-current STM simulations.}
    \label{fig:tbtang_br_au_surface_dbspm_stm}
\end{figure}

\begin{figure}
    \centering
    \includegraphics[width=1\linewidth]{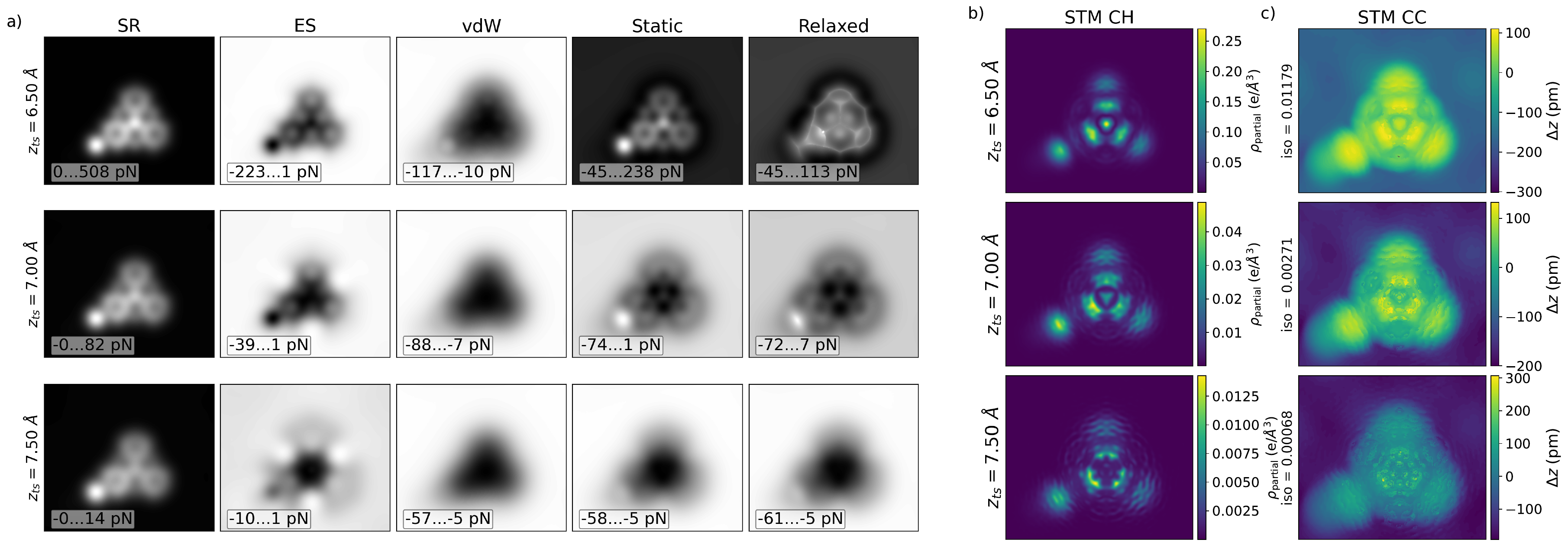}
    \caption{Combined HR-AFM and STM simulations for the bromine-terminated P$^2$TANG edge unit with a Br atom nearby (Fig.~\ref{fig:alt_term_geometries}c). (a) Full density-based AFM (FDBM) simulations showing the short-range (SR), electrostatic (ES), van der Waals (vdW), static, and relaxed contributions at several tip heights with respect to the Au(111) surface. (b, c) Corresponding constant-height and constant-current STM simulations.}
    \label{fig:tbtang_br_br_surface_dbspm_stm}
\end{figure}

\begin{figure}
    \centering
    \includegraphics[width=1\linewidth]{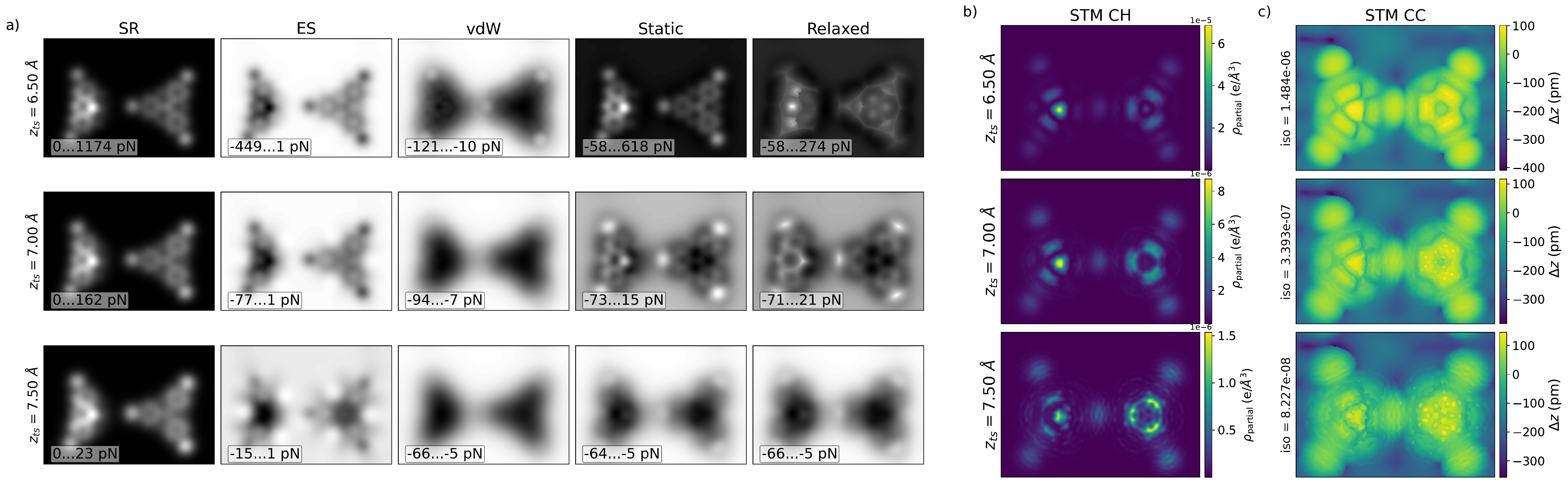}
    \caption{Combined HR-AFM and STM simulations for a putative Br-mediated TANG--Br--TANG junction on Au(111), considered as an alternative assignment for the protrusion observed at the organometallic intermediate in Figure~2. Upon structural relaxation, the bridging Br atom is displaced upward from the Au(111) surface, producing a strongly nonplanar junction. The relaxation sequence leading to this geometry is shown in Figure~\ref{fig:tang_br_tang_relaxation}.(a) Full density-based AFM (FDBM) simulations showing the short-range (SR), electrostatic (ES), van der Waals (vdW), static, and relaxed contributions at several tip heights with respect to the Au(111) surface. (b,c) Corresponding constant-height and constant-current STM simulations. The elevated Br atom produces the characteristic sigma-hole HR-AFM contrast at the center of the junction, unlike the weak contrast observed experimentally at the corresponding intermolecular position, while STM contrast becomes asymmetric. This comparison disfavors a Br-mediated junction as the origin of the experimental STM protrusion.}
    \label{fig:tang-br-tang_afm_stm}
\end{figure}

\begin{figure}
    \centering
    \includegraphics[width=1\linewidth]{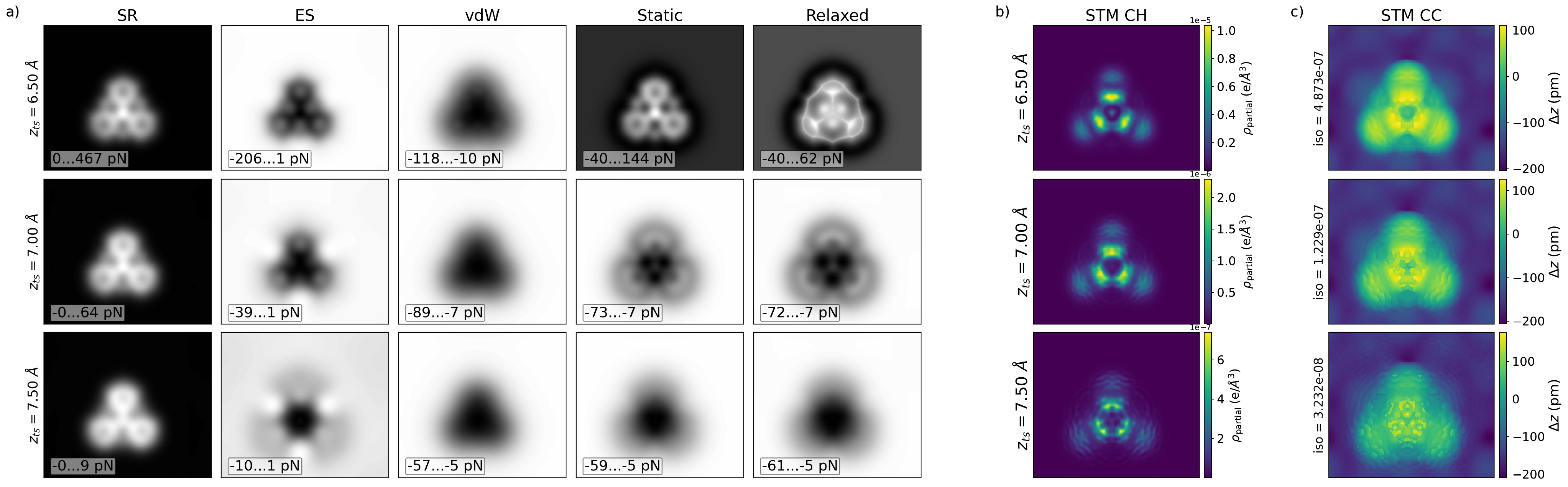}
    \caption{Combined HR-AFM and STM simulations for an H-terminated TANG unit on Au(111), considered as a possible model for the featureless polymer edge terminations. (a) Full density-based AFM (FDBM) simulations showing the short-range (SR), electrostatic (ES), van der Waals (vdW), static, and relaxed contributions at several tip heights with respect to the Au(111) surface. (b,c) Corresponding constant-height and constant-current STM simulations. The hydrogen termination does not introduce an additional prominent terminal feature, illustrating that H passivation is compatible with a comparatively featureless edge contrast, although the present simulations alone do not establish its experimental occurrence.}
    \label{fig:tang_h_terminated_afm_stm_}
\end{figure}

\clearpage

\section{Bending angle clarification}

\begin{figure}
    \centering
    \includegraphics[width=0.8\linewidth]{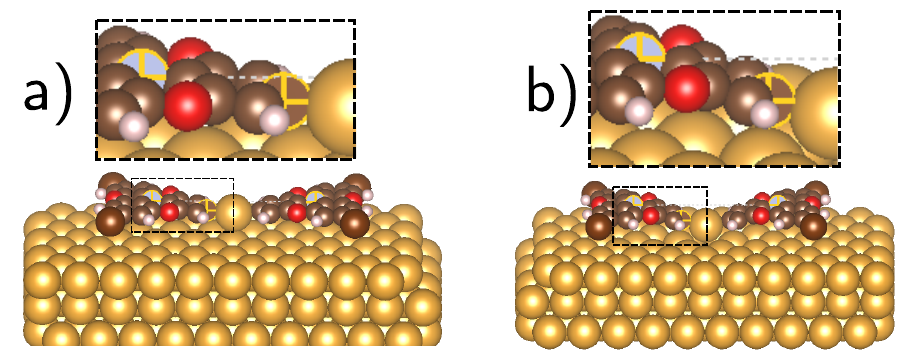}
    \caption{Definition of the molecular bending angle for the partially brominated TANG dimer geometries coordinated to gold. Side views are shown for (a) the Au-adatom coordination geometry and (b) the Au surface-atom coordination geometry. The bending was quantified from the N$'$--N--C angle, where C is the dehalogenated carbon atom bound to Au and the two N atoms define the molecular axis. The Au surface-atom geometry exhibits the larger downward bending toward \ce{Au}(111), with a maximum angle of 18.7~\degree, whereas the Au-adatom geometry shows a reduced bending angle of 9.7~\degree.}
    \label{fig:angle_clarification}
\end{figure}
\clearpage

\section{Structural relaxation of a Br-mediated TANG--Br--TANG junction}

To assess whether residual bromine could account for the characteristic protrusion
observed at the intermolecular junction assigned to the OM2 intermediate, we explored
an alternative Br-mediated configuration. A Br atom was initially placed at a hollow
site of the Au(111) surface between two dehalogenated TANG units, with the two reactive
carbon atoms oriented toward the same Br atom, thereby forming a putative
TANG--Br--TANG junction. This construction was intended to test whether a
surface-bound Br atom could provide a compact intermolecular connection while
simultaneously producing the central STM protrusion observed experimentally.

\begin{figure}
    \centering
    \includegraphics[width=\linewidth]{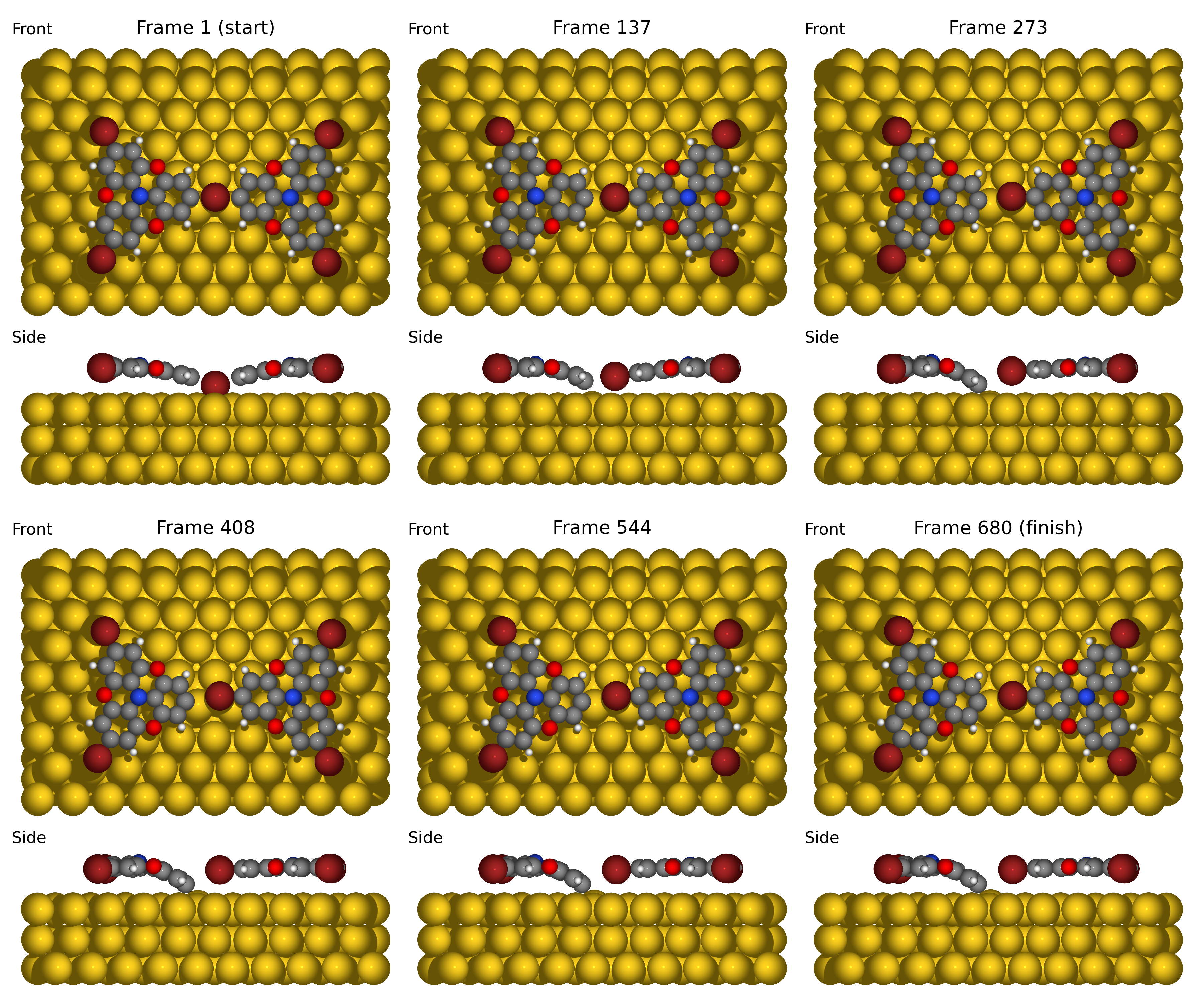}
    \caption{Representative snapshots of the structural relaxation of a putative
    Br-mediated TANG--Br--TANG junction on Au(111). Top and side views are shown for
    the initial geometry (frame 1), intermediate stages of the relaxation
    (frames 137, 273, 408, and 544), and the final relaxed geometry (frame 680).
    During relaxation, the central Br atom is progressively displaced away from the
    Au(111) surface, accompanied by deformation of the adjacent TANG units. The final
    configuration therefore does not preserve the initially constructed, low-lying
    Br-mediated junction. These calculations indicate that such a structure is
    geometrically incompatible with the approximately planar intermolecular junction
    observed experimentally and disfavour residual Br as the origin of the STM
    protrusion assigned to the OM2 intermediate.}
    \label{fig:tang_br_tang_relaxation}
\end{figure}

The resulting structure was then relaxed using DFT. During the relaxation, the
central Br atom is progressively displaced away from its initial hollow-site position
and lifted above the Au(111) surface, accompanied by a distortion of the neighboring
TANG units. The final relaxed structure therefore does not preserve the initially
constructed low-lying Br-mediated junction, instead the junction breaks into two TANG monomers, one coordinated with a Au surface atom and thus highly corrugated and another terminated in bromine that becomes planar. Representative snapshots of the relaxation are shown in
Figure~\ref{fig:tang_br_tang_relaxation}. The corresponding simulated HR-AFM and STM
contrast is shown in Figure~\ref{fig:tang-br-tang_afm_stm}.

\clearpage

\bibliography{bibliography}